# Rationalizing dynamic choices[*]

Henrique de Oliveira     Rohit Lamba

March 2025

**Abstract**

An analyst observes an agent take a sequence of actions. The analyst does not have access to the agent's information and ponders whether the observed actions could be justified through a rational Bayesian model with a known utility function. We show that the observed actions cannot be justified if and only if there is a single deviation argument that leaves the agent better off, regardless of the information. The result is then extended to allow for distributions over possible action sequences . Four applications are presented: monotonicity of rationalization with risk aversion, a potential rejection of the Bayesian model with observable data, feasible outcomes in dynamic information design, and partial identification of preferences without assumptions on information.

## 1 Introduction

As information arrives over time, people may take actions that seemingly go against their own past choices. How can we judge someone's sequence of choices without knowing what they knew? A permissive criterion would allow for *any* sequence of choices that can be explained by the piecemeal arrival of *some* information. The purpose of this paper is to characterize, for a general decision problem, the sequences of choices which can be rationalized by such criteria.

We consider the following model: There is a set of states of the world $\Omega$. The agent, looking to maximize expected utility, starts with a prior $p \in \Delta(\Omega)$, sees a signal $s_1$ that provides her some information about the actual state of the world, and then chooses an action $a_1$. The agent then sees another informative signal $s_2$, chooses an action $a_2$ and so on, until the final action $a_T$ is chosen. A terminal payoff is realized, represented by an arbitrary function $u : A \times \Omega \to \mathbb{R}$, where $A$ is the set of all action sequences.

[*]de Oliveira: Fundação Getúlio Vargas-EESP, henrique.deoliveira@fgv.br; Lamba: Cornell University, rohit-lamba@cornell.edu. We are grateful to Gabriel Carroll (editor) and four anonymous referees for their advice, as well as Ashwin Kambhampati, Tiago Botelho, and Nicolas Goulart for their excellent research assistance. We are also indebted to Dilip Abreu, Nageeb Ali, Pierpaolo Battigalli, Alex Bloedel, Benjamin Brooks, Andrew Caplin, Sylvain Chassang, Laura Doval, Daniel Grodzicki, Danilo de Paula, Vijay Krishna, Jay Lu, Stephen Morris, David Pearce, Luciano Pomatto, Doron Ravid, Larry Samuelson, Ran Shorrer, Ron Siegel, Marciano Siniscalchi, Alexander Wolitzky, and seminar participants at University of Chicago, Yale University, University of Rochester, Indian Statistical Institute Delhi, University of Pennsylvania, New York University, University of British Columbia, Princeton University, Pennsylvania Economic Theory Conference and European Summer Symposium in Economic Theory at Gerzensee for their comments.

Our aim is to characterize the empirical content of this model. To that end, we take the perspective of an outside analyst who knows the mapping $u$, but does not know the agent's prior $p \in \Delta(\Omega)$ nor her information process $\pi : \Omega \to \Delta(S)$, where $S$ is the set of all signal sequences, assumed throughout to be independent of $A$.[1] Upon observing some data about the agent's choices, the analyst asks: Could these data be generated by optimal Bayesian behavior for some $p$ and $\pi$? Put differently, by characterizing the model's empirical content, we mean identifying the restrictions imposed by the joint hypothesis of Bayesian rationality and the specific payoff function $u$, but without any hypothesis on the information structure.

The definition of empirical content depends on what data the analyst can observe. We start with the parsimonious assumption that the analyst observes a single action sequence $(a_1, \ldots, a_T)$. Such an action sequence can then be rationalized by an analyst if there is a $p$ and $\pi$ and some optimal strategy that chooses the action sequence with positive probability.

To understand the setting and what this definition allows, consider the following example:

**Example 1.** A CEO faces an opportunity to invest in a project with uncertain payoffs: there is a return of 4 if the project meets favorable conditions in the future (good state) and 0 if not (bad state). The project bears fruits on two rounds of investment, and each round of investment costs 1 unit. The CEO has three options: not invest, invest in the first round and pull back in the second, or invest in both periods. The CEO's payoff matrix and decision tree can be summarized as follows:

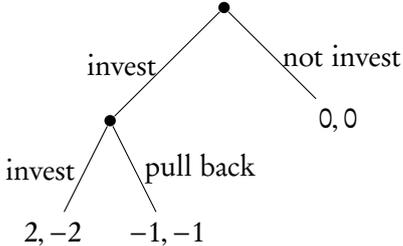

Suppose we learn that the CEO invested in the first round, incurring the initial cost, but then pulled back. Some might interpret that as evidence of incompetence, saying that in no state can this sequence of actions be justified. They might say that even if the CEO were not sure about the state of the world, not investing would surely have been a better choice. These critics would be ignoring a simple explanation: it might be that the CEO initially received good news about the investment, but after the first round of investment, learnt that the project was likely to fail.

In Example 1, the action sequence (*invest*, *pull back*) is what we call *apparently dominated*—there exists another sequence of actions, (*not invest*, ∅), under which the agent does strictly better in every state of the world.[2] It will be easy to show that any action sequence that cannot be rationalized is apparently dominated. However, as Example 1 shows, the converse is false. In fact, in Example 1, all three possible sequences of choices can be rationalized, illustrating how permissive this first

---

[1] The assumption that the analyst knows $u : A \times \Omega \to \mathbb{R}$ is quite flexible. We can alternatively think of a family of utility functions $u_{\omega,\lambda} : A \to \mathbb{R}$, where $\omega$ indexes a class of utility functions that the agent may learn over time, and $\lambda$ is a parameter that the agent knows but the analyst does not (see Example 2).

[2] Generally, an action sequence is apparently dominated if another action sequence (or a lottery over action sequences) does strictly better in every state of the world.

criterion is. But it is not vacuous and can exclude some dynamic choices. For instance, consider the following example:

**Example 2.** A firm can bet on one of two technologies, $X$ or $Y$. The firm can also postpone the decision, but by doing so, its payoff is discounted by a factor $\delta$, where $0 < \delta < 1$. The payoff matrix and decision tree are as follows:

|   | $x$ | $y$ | $wx$ | $wy$ |
|---|---|---|---|---|
| X | 5 | 3 | $5\delta$ | $3\delta$ |
| Y | 3 | 5 | $3\delta$ | $5\delta$ |

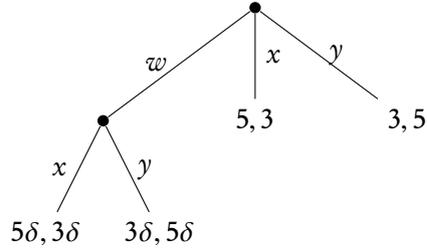

Note that both $wx$ and $wy$ are apparently dominated, which does not necessarily rule them out.[3] We learn that the firm has decided to wait instead of making an immediate bet. Under what values of $\delta$ can this choice be rationalized? By waiting, the firm can get at most $5\delta$. By making an immediate decision, the firm is guaranteed to get at least 3. Hence, if $\delta < {}^3\!/\!_5$, waiting cannot be rationalized.

But this is not the whole story. If the firm makes an immediate decision to randomize equally between $x$ and $y$, it is guaranteed an expected payoff of 4, no matter the state. Therefore, waiting cannot be rationalized when $\delta < {}^4\!/\!_5$. On the other hand, if $\delta \geq {}^4\!/\!_5$, waiting can be explained by the following information: it could be that the firm starts with an even prior and then fully learns the state of the world in the second period. Thus, waiting can be rationalized precisely when $\delta \geq {}^4\!/\!_5$.

More generally, an action sequence can be rationalized when there exists a prior $p$ and an information structure $\pi$ for which an optimizing agent could choose that action sequence with positive probability. Thus, to argue that an action sequence can be rationalized, it is enough to provide a single information structure and prior that proves it to be so; to argue that an action sequence cannot be rationalized, it must be shown that every information structure and prior would fail to rationalize it. In Example 2, we found a single deviation that simultaneously showed that every information structure would fail to rationalize waiting, thereby avoiding directly considering the set of all information structures.

The challenge now is: for any arbitrary set of states, actions, and utility function, in order to show that an action sequence cannot be rationalized, can we generalize the deviation argument? The construction of this argument through a *deviation rule* forms the core of our paper.

Formally, a deviation rule is an *adapted* mapping from action sequences to lotteries over action sequences, $D : A \to \Delta(A)$. Adaptedness simply requires that deviations today can only be a function of past actions and past deviations, not future actions or deviations. In Example 1, if we map (*invest*, *pull back*) to (*not invest*, ∅), then adaptedness demands that we have to map (*invest*, *invest*) also to (*not invest*, ∅). As a result, an action sequence and a state of the world exist for which the deviation makes the agent worse off. In Example 2, the (perhaps intuitively appealing) mapping $wx \mapsto x$, $wy \mapsto y$, $x \mapsto x$ and $y \mapsto y$ is not adapted, and hence not a valid deviation rule. However, the mapping $wx \mapsto \frac{1}{2}x + \frac{1}{2}y$, $wy \mapsto \frac{1}{2}x + \frac{1}{2}y$, $x \mapsto y$ and $y \mapsto y$ is adapted, and is eventually used to show that action sequences $wx$ and $wy$ cannot be rationalized if $\delta < 4/5$.

---
[3]We are using the shorthand $wx$ for $(w, x)$ and $wy$ for $(w, y)$.

The concept of deviation rules is bereft of information since it must respect the constraint that the analyst may not know anything about the agent's sequential information structure. For any strategy of the agent $\sigma : S \to \Delta(A)$, the composition mapping $D \circ \sigma = \sigma'$ is a new strategy that tells the agent the following: upon observing a sequence of signals $s = (s_1, ..., s_T)$, if $\sigma$ had generated a distribution $\sigma(s) \in \Delta(A)$, now instead generate the distribution $\sigma'(s) \in \Delta(A)$. Since $\sigma$ and $D$ are both adapted, the new strategy $\sigma'$ is also adapted and hence well defined.

We say that a deviation rule improves upon an action sequence if, for every state of the world, it strictly increases payoffs along that action sequence without worsening payoffs elsewhere on the decision tree. We then say that the action sequence is *truly dominated* by this proposed deviation rule. In Example 1, the action sequence (*invest*, *pull back*) cannot be improved upon without worsening payoffs along the rest of the decision tree and hence is not truly dominated. On the other hand, in Example 2 the action sequences $wx$ and $wy$ can be improved upon by the deviation rule described above without tinkering with payoffs elsewhere and hence are truly dominated. Our main result, Theorem 1, establishes the following equivalence: An action sequence cannot be rationalized if and only if it is truly dominated.

The theorem can be viewed as a form of duality—it replaces the "for all" quantifier with the "there exists" quantifier and vice-versa. To show that an action sequence can be rationalized, the analyst can construct *one* information structure for which the action sequence receives positive weight under an optimal strategy. To show that an action sequence cannot be rationalized, the analyst can now construct *one* deviation rule that dominates it.[4]

By delineating the sequences of actions that cannot be rationalized, Theorem 1 settles the question of empirical content when a single action sequence is observed. This is a minimal data requirement that allows us to make predictions even for an individual agent. But this minimality means that the theory may not be rejected. In some cases, such as Example 1, the theory cannot be rejected by any observation. It is then natural to look for finer predictions coming from richer datasets.

Theorems 2 and 3 concern data in the form of entire distributions, such as what can be obtained from a large sample of identical agents with independent information. In this context, Theorem 1 pins down the support of any distribution of action sequences that an analyst could observe in the population. Now, observing the choices of an entire population, can the analyst go further and ask which distributions can be rationalized?

---

[4] Note that if $T = 1$, the actions which can be rationalized are precisely those that are a best-response to some belief over states. The theorem then reduces to the celebrated Wald-Pearce Lemma (Wald [1949] and Pearce [1984]), which states that the actions that are never a best-response, and hence cannot be rationalized, are strictly dominated by some mixed strategy. Here, our rule would recommend deviating from the dominated action to the dominating mixed strategy and not deviating from the other actions. Adaptedness, of course, has no bite in the static model.

More concretely, in Example 1, we argued that the action sequence (*invest*, *pull back*) can be rationalized. It is easy to see that it cannot be rationalized with probability one, that is, no sequential information structure can induce a rational firm to take this action sequence for sure; it might as well choose (*not invest*, ∅) no matter the signals it observed. The population interpretation is that there must be an upper bound on the fraction of firms that choose (*invest*, *pull back*) in the data set. Similarly, in Example 2, what is the upper bound on the fraction of agents that choose $wx$ or $wy$ as a function of $\delta$? The next two results help the reader answer these questions by characterizing the family of distributions over action sequences that can be rationalized.

Suppose the analyst observes chosen action sequences along with the associated realized states for a large number of decision problems. For Example 1, this data requirement would collate entries in one of six possible bins:

| (*not invest*, ∅), good | (*invest*, *pull back*), good | (*invest*, *invest*), good |
|---|---|---|
| (*not invest*, ∅), bad | (*invest*, *pull back*), bad | (*invest*, *invest*), bad |

The objective now is to explain when a joint distribution $\gamma \in \Delta(A \times \Omega)$ can be rationalized. For its dual counterpart, we say that a deviation rule dominates $\gamma$ if it generates a strict improvement in expected payoffs. We call this *average dominance*, where the expectation over actions and states is taken for the distribution generated by the composition map between the deviation rule and whatever strategy a representative agent follows. Theorem 2 then provides the dual characterization that a joint distribution cannot be rationalized if and only if it is averagely dominated. This result boils down to a set of inequalities corresponding to obedience constraints familiar from information design (see Bergemann and Morris [2016]), extended here to a dynamic environment.

In some scenarios, the analyst may only observe the set of action sequences but not the realized states. So, in the context of Example 1, the analyst records which of the three possible action sequences were chosen by each firm in the data set, but does not know what were the underlying fundamentals associated with each of those decisions. The objective now is to explain when a marginal distribution $\bar{\gamma} \in \Delta(A)$ can be rationalized. The notion of dominance that pins down the duality is more nuanced. It combines ideas on dominance used for Theorems 1 and 2, resulting in an *intermediate* notion of *dominance*. It considers deviation rules that take the worst-case improvement over states for each action sequence, and then averages these values over action sequences using $\bar{\gamma}$. Theorem 3 states that a distribution $\bar{\gamma}$ cannot be rationalized if and only if it is intermediately dominated.

Four applications are presented in Section 8: First, by characterizing the empirical content of the model, the three theorems provide a clear method for testing for Bayesian rationality. For Example 1, using Theorem 3, we can conclude that $2/3$ is the upper bound on the fraction of firms that can rationally be seen making the choice (*invest*, *pull back*). Thus, if more than two-thirds of firms in the data set facing Example 1 choose (*invest*, *pull back*), the Bayesian model is rejected. Similarly, for Example 2, we show that the maximal fraction of agents that can rationally be seen to choose to wait in the first period is given by 0 when $\delta < 4/5$ and by $3 - 2/\delta$ when $\delta \geq 4/5$, which converges to 1 as $\delta$ converges to 1.[5]

Second, if one *assumes* Bayesian rationality, the results can be used to partially identify parameters from the agent's preferences, without imposing assumptions on information. For instance, the firm's choice to wait in Example 2 helps identify the cost of waiting to be $\delta \geq \frac{4}{5}$. More generally, for a population of firms, we show that a distribution that puts weight $\gamma^w > 0$ on $wx$ (or $wy$) generates a lower bound given by $\delta \geq \max\left\{\frac{2}{3-\gamma^w}, \frac{4}{5}\right\}$. This insight is formalized more generally in a partial identification result that speaks to recent work in applied econometrics.

Next, besides these practical applications, our framework can also be used conceptually. As an illustration, we show that the set of action sequences that can be rationalized is an increasing function of risk aversion—the more risk averse the agent, the harder it is to rule out action sequences. This generalizes a similar observation by Weinstein [2016] and Battigalli, Cerreia-Vioglio, Maccheroni, and Marinacci [2016] from static to dynamic. Deviation rules are again critical in the argument, and it is unclear how to prove such a result without them.

Finally, deviation rules allow us to write dynamic obedience constraints, providing a dual method to solve dynamic information design problems. This is illustrated through a simpler version of the *moving the goalposts* model studied by Ely and Szydlowski [2020], where the agent is induced to take an apparently dominated action sequence with maximal possible probability.

The logic of deviation rules is applicable beyond the framework presented here. After the initial draft of our paper was circulated, Makris and Renou [2023] has shown how the notion of true dominance can be expanded, and thus Theorem 1 can be modified to allow for sequential information structures that depend on actions; similar modifications for Theorems 2 and 3 remain open questions. We are hopeful that a wider set of applications will emerge in decision theory, applied econometrics, behavioral economics, and information and mechanism design, some of which are described in Section 8.

## 2 Model and definitions

### 2.1 Notation

A *stochastic map* from $X$ to a finite set $Y$ is a function $\alpha : X \to \Delta(Y)$, where $\Delta(Y)$ is the set of probability distributions over $Y$. We represent the probability assigned to $y$ at the point $x$ by $\alpha(y|x)$. The composition of two stochastic maps $\alpha : X \to \Delta(Y)$ and $\beta : Y \to \Delta(Z)$ is defined by

$$\beta \circ \alpha(z|x) = \sum_{y \in Y} \beta(z|y)\alpha(y|x).$$

---

[5] There are two key steps here: the construction of the appropriate deviation rule for values greater than the upper bound and a rationalizing information structure for values below the upper bound.

We can think of a lottery as a stochastic mapping whose domain is a singleton. Therefore, given $\alpha \in \Delta(Y)$ and $\beta : Y \to \Delta(Z)$, we write

$$\beta \circ \alpha(z) = \sum_{y \in Y} \beta(z|y)\alpha(y)$$

to be the probability with which $z$ is chosen by $\beta \circ \alpha$.

For a real-valued function $u : Y \to \mathbb{R}$ and for a lottery $\alpha \in \Delta(Y)$, we denote by $u(\alpha) = \sum_{y \in Y} \alpha(y)u(y)$ the expected value of $u(\cdot)$ under the distribution $\alpha$.

Throughout the text, we consider a finite number of time periods $t = 1, \ldots, T$. For a collection of sets $(X_t)_{t=1}^T$, we will use the following notation

$$X^t = \prod_{\tau=1}^t X_\tau \qquad X = \prod_{\tau=1}^T X_\tau$$

with elements $\mathbf{x}^t \in X^t$ and $\mathbf{x} \in X$. Finally, a stochastic map $\alpha : X \to \Delta(Y)$ is said to be *adapted* if the marginal probability of the first $t$ terms of $\mathbf{y}$ depends only on the first $t$ terms of $\mathbf{x}$; formally, it is adapted if the function

$$\sum_{y_{t+1},\ldots,y_T} \alpha(y_1,\ldots,y_t,y_{t+1},\ldots y_T|x_1,\ldots,x_t,x_{t+1},\ldots,x_T)$$

is constant in $x_{t+1}, \ldots, x_T$.

## 2.2 The model

In each time period $t$, the agent chooses an action $a_t$ from a finite set $A_t$. Payoffs are determined after period $T$ by a utility function $u(\mathbf{a}, \omega)$, which depends on the entire action sequence $\mathbf{a} = (a_1, \ldots, a_T) \in A$ and a potentially unknown state of the world $\omega$ drawn from a finite set $\Omega$. There are no other restrictions on the utility function.

The agent is informed about the underlying state of the world over time through a sequence of signals. The timeline of the dynamic decision problem is expressed in Figure 1. Every period, before taking an action, the agent observes a signal that is (potentially) correlated with the state of the world and with the signals she has observed in the past. Formally, the sequence of signals is generated by a *sequential information structure*:

**Definition 1.** *A **sequential information structure** is a sequence of finite sets of signals $(S_t)_{t=1}^T$ and a stochastic mapping $\pi : \Omega \to \Delta(S)$.*[6]

---

[6]We can equivalently define the sequential information structure period-by-period as follows. Let $\pi = (\pi_t)_{t=1}^T$ be a family of stochastic mappings where $\pi_1 : \Omega \to \Delta(S_1)$, and $\pi_t : \Omega \times S^{t-1} \to \Delta(S_t) \ \forall \ 2 \leq t \leq T$. Except for zero probability events, we can deduce that the two definitions are equivalent. The minor distinction does not affect the agent's utility and is therefore irrelevant for our results. For a proof, see Lemma 3 in de Oliveira [2018]. See also the definition of "sequential decision procedure" in Greenshtein [1996].

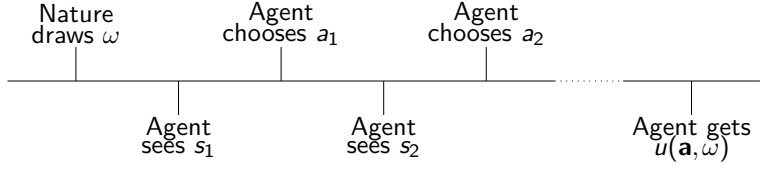

Figure 1: The timeline of signals and actions

We will often refer to the sequential information structure simply as $\pi$; the set of signals shall be implicit.[7] The agent's strategy maps each sequence of signals into a lottery over actions every period, with the restriction that the agent cannot base the choice of an action on signals that have not yet been revealed, which we call adaptedness.

**Definition 2.** *A* **strategy** *for the agent is an adapted stochastic mapping $\sigma : S \to \Delta(A)$.*[8]

Given the sequential information structure $\pi$ and agent's strategy $\sigma$, the probability that the agent takes a given sequence of actions in each state of the world $\omega$ is given by $\sigma \circ \pi(\mathbf{a}|\omega)$. Finally, given a prior $p \in \Delta(\Omega)$, she can evaluate her expected payoff:

$$U(\sigma, \pi, p) = \sum_{\omega \in \Omega} p(\omega) \sum_{\mathbf{a} \in A} \sigma \circ \pi(\mathbf{a}|\omega) u(\mathbf{a}, \omega).$$

The agent's problem then is to choose an optimal $\sigma$ given $\pi$ and $p$. Throughout the paper, we refer to this model of decision-making as the *Bayesian model*.

Our goal is to characterize the empirical content of this model: identifying the restrictions imposed by the joint hypothesis of Bayesian rationality and the specific payoff function $u$, but without any hypothesis on the information structure. To that end, we say that an action sequence can be *rationalized* if it can be chosen with positive probability by an optimizing agent with some information structure and some prior.

**Definition 3.** *An action sequence $\mathbf{a} \in A$ can be* **rationalized** *if there exists a triplet $(\sigma, \pi, p)$ such that:*

1. $\sigma \in \arg\max_{\hat{\sigma}} U(\hat{\sigma}, \pi, p)$ *and*

2. $\sigma \circ \pi \circ p(\mathbf{a}) > 0$.[9]

This definition is permissive in the sense that an action sequence is considered rationalized even if its probability is very small, so long as it is positive. Moreover, because the agent sees a signal before choosing the first action, any two interior prior beliefs $p$ and $p'$ result in the same criterion, since we can always consider a signal distribution which updates from $p$ to $p'$ with positive probability. In that sense, the choice of prior, in addition to the choice of the sequential information structure, arms the analyst with more instruments than she requires to rationalize an action sequence. However, fixing a prior that puts zero probability on some states loses generality, since updated beliefs must also put zero probability on those states.[10]

---

[7] We restrict attention to exogenous learning, that is, information structures that do not depend on the actions chosen by the agent.

[8] As with information structures, an equivalent way to think of the agent's strategy is a family of stochastic mappings $\sigma = (\sigma_t)_{t=1}^{T}$, where $\sigma_1 : S_1 \to \Delta(A_1)$, and $\sigma_t : S^t \times A^{t-1} \to \Delta(A_t) \ \forall \ 2 \leq t \leq T$. It is possible to deduce one formulation from the other.

[9] Here $\sigma \circ \pi \circ p(\mathbf{a}) = \sum_\omega \sigma \circ \pi(\mathbf{a}|\omega) p(\omega)$ (see Section 2.1).

To deduce that an action sequence cannot be rationalized, the analyst needs to work through all possible pairs $(\pi, p)$, and show that the corresponding optimal strategy $\sigma$ will not pick that action sequence with positive probability. Since the set of all sequential information structures is quite large, this poses a challenge. Our main goal is to find an alternative way to characterize the set of action sequences that cannot be rationalized.

A final aspect of the model is the knowledge set of the outside analyst. Of course, the analyst observes the action sequence in Definition 3. In addition, it is assumed that the analyst knows some aspect of the mapping $u : A \times \Omega \to \mathbb{R}$. A simple way to understand the permissiveness of the claim is to rewrite the utility function as $u_{\omega,\lambda} : A \to \mathbb{R}$, where $\omega \in \Omega$ can embed a class of utility functions that the agent learns over time, and $\lambda$ in some compact set, as we saw in Example 2 with the discount factor $\delta$, is a parameter to be estimated that the agent knows but the analyst doesn't (see also Section 8.4) . This incorporates two subtleties: first, that the model can accommodate a large class of possible utilities through the set $\Omega$, and second, the utility functions can accommodate missing parameters, which the analyst must elicit from the choice data. Of course, for the model to generate some empirical content, we have to impose some structure on preferences, for we are not imposing any structure on information.

## 3 The static problem

To fix ideas, it is easiest to start from the simple case of $T = 1$. In this static problem, the agent starts with a prior $p$, observes a signal $s$, and takes an action $a$, resulting in a payoff $u(a, \omega)$. Letting

$$q(\omega|s) = \frac{\pi(s|\omega)p(\omega)}{\pi \circ p(s)}$$

denote the posterior belief of the agent upon seeing $s$, we can rewrite the agent's expected utility from choosing strategy $\sigma$ as:

$$U(\sigma, \pi, p) = \sum_{\omega,a,s} u(a,\omega)\sigma(a|s)\pi(s|\omega)p(\omega) = \sum_{\omega,a,s} u(a,\omega)\sigma(a|s)q(\omega|s)\pi \circ p(s). \quad (1)$$

---

[10]This logic can be pushed further: to determine the set of actions that can be rationalized going forward, the only relevant aspect of a belief is the set of states that have zero probability. So, a behavioral model where agents may violate the martingale condition of beliefs could rationalize the same set of action sequences as the Bayesian model, as long as its belief process agrees with the Bayesian belief process on which states have zero probability. We are grateful to Andrew Caplin for pointing this out to us.

This makes the agent's problem separable in $s$, so it reads: for each $s$, choose an action $a$ to maximize

$$\sum_{\omega \in \Omega} u(a,\omega) q(\omega|s). \qquad (2)$$

Therefore, an action can be rationalized if and only if it is a *best-response to some posterior belief $q$*. Hence, to find if an action can be rationalized, we can restrict attention to the case where the agent starts with a "prior $q$" and learns nothing thereafter. In particular, if an action can be rationalized, there is a triplet $(\sigma, \pi, p)$ where $\sigma$ is optimal and chooses that action with probability 1. To summarize:

**Remark 1.** *Let $T = 1$. Then the following statements are equivalent:*

1. *There exists $q$ such that $a$ maximizes (2);*

2. *There exists $(\sigma, \pi, p)$ such that $\sigma$ maximizes (1) with $\sigma(a) > 0$;*

3. *There exists $(\sigma, \pi, p)$ such that $\sigma$ maximizes (1) with $\sigma(a) = 1$.*

An elegant duality result by Wald [1949] and Pearce [1984] characterizes what it means for an action to be rationalized in the static model.[11] The result states that, in a two-player game,

**Lemma 1** (Wald-Pearce). *An action is never a best response if and only if it is strictly dominated by some mixed strategy.*

In our context, think of a game where Player 1 is our agent, choosing action $a$, and Player 2 is Nature, choosing state $\omega$. A mixed strategy $\alpha \in \Delta(A)$ strictly dominates $a$ if and only if $u(\alpha, \omega) > u(a, \omega)$ for all $\omega \in \Omega$, where $u(\alpha, \omega)$ is the expected utility of following that mixed strategy. An action $a$ is then said to be strictly dominated if there exists a mixed strategy $\alpha$ that strictly dominates it. Given Remark 1, $a$ is never a best response if and only if it cannot be rationalized. Therefore

**Corollary 1.** *For $T = 1$, an action $a$ cannot be rationalized if and only if it is strictly dominated.*

The key idea behind the Wald-Pearce lemma is that it is possible to invert the order of quantifiers in the statement "for all $q \in \Delta(\Omega)$, there exists $\alpha \in \Delta(A)$ such that $\mathbb{E}_q[u(a,\omega)] < \mathbb{E}_q[u(\alpha,\omega)]$". This can be seen, for example, by constructing a zero-sum game where nature picks the belief $q$ and the agent picks an alternative action $\alpha$ (possibly mixed). Using the min-max theorem, we get

$$\min_q \max_\alpha \mathbb{E}_q[u(\alpha,\omega) - u(a,\omega)] = \max_\alpha \min_q \mathbb{E}_q[u(\alpha,\omega) - u(a,\omega)].$$

When $a$ cannot be rationalized, the above expression is positive and bounded away from zero. More specifically, the positivity of the left-hand side is equivalent to $a$ not being rationalized, and the positivity of the right-hand side is equivalent to it being strictly dominated.

---

[11]Wald [1949], in particular, treats the problem in great generality, allowing for infinitely many actions. Whenever we refer to Wald's result in this paper, we mean a simplified version with finitely many actions.

The two theoretical challenges for us are (i) to formulate the right notion of what it means for an action sequence to be dominated in the sequential model, and (ii) to establish the appropriate inversion of quantifiers for our framework. We start by defining the appropriate notion of domination in the sequential model.

## 4 Deviation rules and true dominance

### 4.1 A necessary but not sufficient condition

An obvious notion of dominance that does not rely on information structures is the following: a sequence of actions is "dominated" if there exists another sequence of actions that does strictly better in every state of the world. We will refer to this as *apparent dominance*. Recollect that the payoff from a randomized action sequence $\alpha \in \Delta(A)$ is denoted by $u(\alpha, \omega) = \sum_{\mathbf{a} \in A} \alpha(\mathbf{a}) u(\mathbf{a}, \omega)$, where $\alpha(\mathbf{a})$ refers to the probability of action sequence $\mathbf{a}$ under $\alpha$.

**Definition 4.** *An action sequence $\mathbf{a} \in A$ is* **apparently dominated** *if there exists a randomized action sequence $\alpha \in \Delta(A)$ such that*
$$u(\alpha, \omega) > u(\mathbf{a}, \omega) \ \forall \ \omega \in \Omega.$$

Every action sequence that cannot be rationalized is apparently dominated, making it a necessary condition for our endeavored characterization. That is, if an action sequence is not apparently dominated, we can always find an information structure such that the optimal strategy corresponding to it chooses the action sequence with positive probability. The following proposition formalizes the claim.

**Proposition 1.** *Suppose $\mathbf{a} \in A$ cannot be rationalized. Then, $\mathbf{a}$ must be apparently dominated.*

*Proof.* Suppose $\mathbf{a}$ is not apparently dominated. By Lemma 1, the Wald-Pearce Lemma, $\mathbf{a}$ must be a best-response to some static "belief $p$". Letting $p$ be the prior and $\pi$ be completely uninformative, a best response to $(p, \pi)$ is the strategy that always chooses $\mathbf{a}$. □

Even though apparent dominance is a demanding condition, an apparently dominated action sequence can be rationalized. In Example 1, the action sequence $a_1 = $ *invest* and $a_2 = $ *pull back* is apparently dominated by the action sequence $a_1 = $ *not invest* and $a_2 = \emptyset$. Yet it is easy to construct an information structure where it will be optimal for the agent to choose (*invest*, *pull back*) with positive probability.[12]

Notice that the apparent dominance of (*invest*, *pull back*) can be established simply by comparing its payoffs with that of *not invest*. The payoffs for (*invest*, *invest*) are therefore irrelevant. Yet, when the good state is very likely, these payoffs are precisely what motivates the agent to do the initial investment. When we see that the agent chose (*invest*, *pull back*), the fact that the agent *could* have ended up choosing (*invest*, *invest*) makes those payoffs relevant.

---

[12]The first period signal tells the agent that the good state is highly likely, only to reveal in period two through the second signal that the bad state is now more likely.

Therefore, something more than apparent dominance is required for an action sequence not to be rationalized. In addition to improving upon the action sequence under consideration, that "more" needs to evaluate other sequences of actions that the agent might expect to choose. This motivates the definition of a deviation rule, which prescribes not only how the agent should deviate from the *observed* action sequence, but in *every* possible action sequence.

## 4.2 Deviation rules and true dominance

A *deviation rule* is an adapted mapping $D : A \to \Delta(A)$, where recollect that being adapted means that the marginal distribution on $A^t$, the (potentially random) deviation strategy for the first $t$ periods, depends only on $A^t$, the first $t$ elements of the original strategy from which the agent is deviating. We can think of the deviation rule as a list of alternative actions the agent would take as a function of the actions she originally intended to take. As will be relevant later, you can also think of a deviation rule as a strategy, where the signals are action recommendations. Importantly, a deviation rule is a fully prescribed plan so that if $\sigma : S \to \Delta(A)$ is the original strategy, then $D \circ \sigma : S \to \Delta(A)$ is also a well-defined strategy.

Now, we are in a position to define the appropriate notion of dominance for our model.

**Definition 5.** *A deviation rule $D : A \to \Delta(A)$ **dominates** an action sequence $\mathbf{a} \in A$ if*

1. *$u(D(\mathbf{a}), \omega) > u(\mathbf{a}, \omega)$ for all $\omega \in \Omega$.*

2. *$u(D(\mathbf{b}), \omega) \geq u(\mathbf{b}, \omega)$ for all $\mathbf{b} \in A$ and $\omega \in \Omega$.*

*We say that $\mathbf{a}$ is **truly dominated** if there exists a deviation rule that dominates it.*

The first part of the definition requires that the action sequence to be dominated is strictly upon. The second part requires that the payoff induced by the deviation rule shouldn't become worse for any other action sequence in any state. Moreover, there's no explicit time dimension in the definition above; time is implicit in the condition that $D$ must be adapted. For $T = 1$, the same definition applies, but the condition that $D$ is adapted becomes vacuous, as does the second part of the definition. In that case, if $a$ is strictly dominated by $\alpha$, we can define a deviation rule $D_\alpha$ which takes $a$ to $\alpha$ and does not change any other actions. $D_\alpha$ then dominates $a$ according to the definition above.

When $T > 1$, the adaptedness restriction prevents the construction of such a simple deviation rule—if $D$ specifies a change for the first action in the sequence $\mathbf{a}$, then it must specify the same change for all sequences $\mathbf{b}$ which share that same first action, and so on. The second condition and the embedded notion of adaptedness in the definition impose meaningful restrictions when $T > 1$, encapsulating the distinction between true and apparent dominance.

## 4.3 Discussion

To better grasp the definitions of deviation rule and true dominance, here we illustrate the concepts in the context of our examples. For the decision trees depicted in Figure 2, each complete sequence of actions corresponds to a terminal node. Thus any mapping from sequences of actions into sequences of actions is depicted as arrows between terminal nodes.

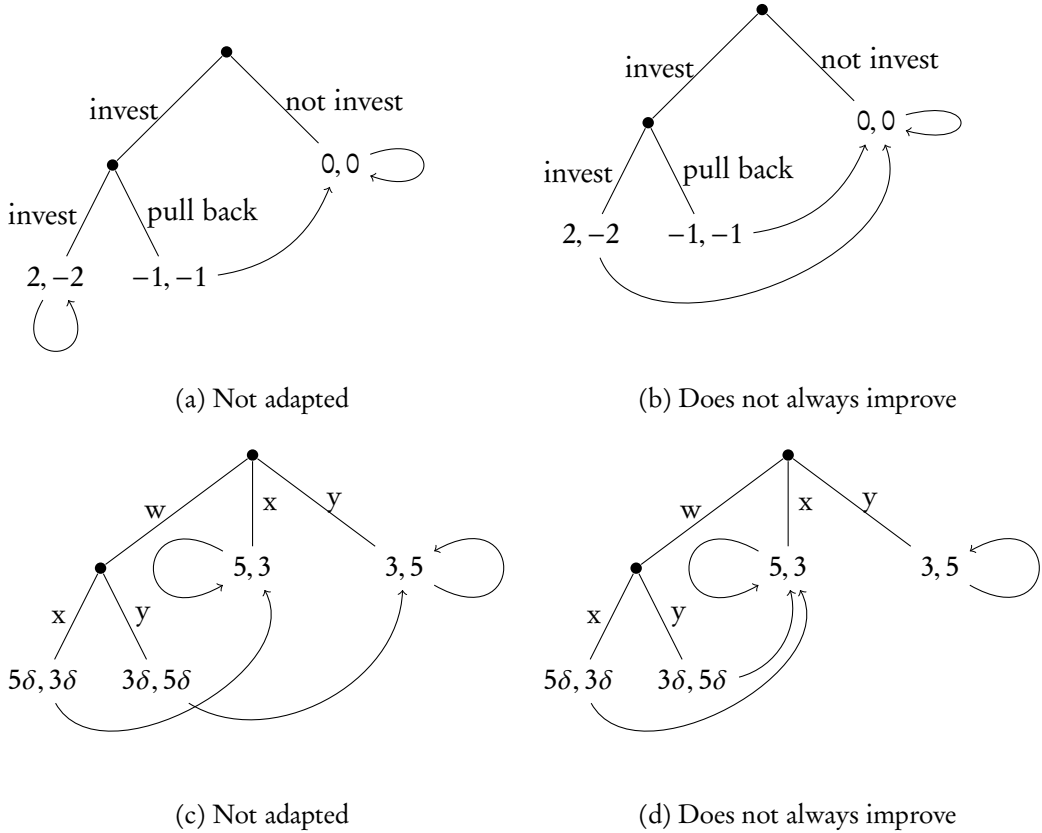

(a) Not adapted

(b) Does not always improve

(c) Not adapted

(d) Does not always improve

Figure 2: Deviation rules for Examples 1 and 2

Figures 2a and 2b depict the decision tree for Example 1. Since the sequence of actions (*invest, pull back*) is apparently dominated by *not invest*, we may try to find a deviation rule that dominates (*invest, pull back*). The simplest such proposal would be that the agent should choose *not invest* whenever she was going to choose (*invest, pull back*), as shown in Figure 2a. However, at the time the agent is choosing to invest, she may not yet know whether she will pull back in the future. The impracticality of this proposal is reflected in the fact that this "deviation rule" is not adapted. If we want the agent to never invest whenever she was going to choose (*invest, pull back*), we must also recommend that she never invest when she was going to choose (*invest, invest*), as in 2b. But although the deviation rule in 2b is now adapted, it worsens payoffs for the action sequence (*invest, invest*) in the good state; thus, it violates part 2 of Definition 5.

Similarly, in the waiting example, the "deviation rule" depicted in Figure 2c is not adapted, since it represents the infeasible advice "whatever you would choose in the second period, choose the same in the first period". The deviation rule in Figure 2d represents the advice "if you were thinking about waiting, choose $x$ instead", which is adapted. When $\delta < \frac{3}{5}$, it dominates $wx$ and $wy$, but when $\delta > \frac{3}{5}$ it does not dominate $wx$ nor $wy$, because $x$ may give a strictly lower payoff than $wy$. For the tightest possible statement, we therefore constructed the deviation rule $wx \mapsto \frac{1}{2}x + \frac{1}{2}y$, $wy \mapsto \frac{1}{2}x + \frac{1}{2}y$, $x \mapsto y$ and $y \mapsto y$ which (simultaneously) truly dominates $wx$ and $wy$ if and only if $\delta < \frac{4}{5}$.

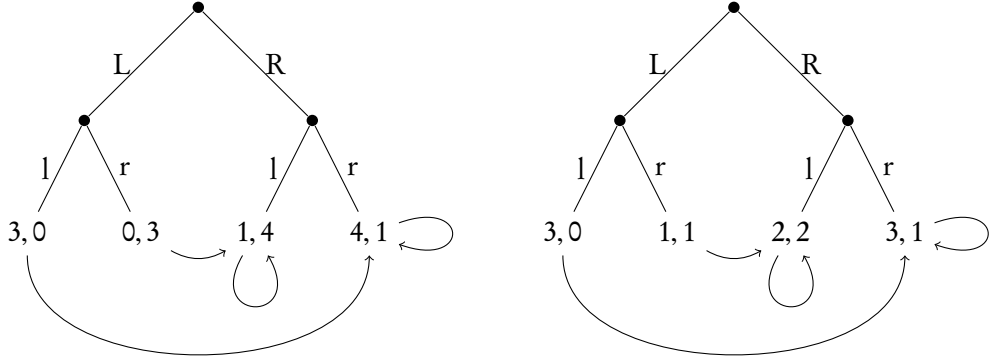

(a) Adapted and improves upon Ll & Lr    (b) Adapted and improves upon Lr

Figure 3: Deviation rules with history dependence

Our examples so far have featured simple first-period deviations. Figure 3 shows how history-dependent deviations may be required to establish that an action sequence is truly dominated. In Figure 3a, both $(L, l)$ and $(L, r)$ are truly dominated by the deviation rule depicted. Despite having the same first period action, $(L, l)$ and $(L, r)$ are deviated to different action sequences: $(R, r)$ and $(R, l)$, respectively; hence the history dependence in deviations. Analogously, $(Lr)$ is shown to be truly dominated in 3b by the same deviation rule, but note that here $(L, l)$ is not truly dominated.

## 5 The main result

We now state our main result.

**Theorem 1.** *A sequence of actions cannot be rationalized if and only if it is truly dominated.*

The theorem provides a tight characterization of the set of action sequences that cannot be rationalized. Through its duality formulation, it simplifies their identification by requiring the analyst to construct *one* deviation rule as opposed to treading through the family of *all* sequential information structures.

The steps involved in establishing Theorem 1 are divided into three subsections. First, we state the *Obedience Principle*: any sequential information structure is equivalent to a canonical information structure, wherein at each point in time the agent is recommended an action which is in her interest to follow. Second, we state a *switching lemma* that will allow us to invert the order of quantifiers. Third, we prove the result.

### 5.1 Obedience principle

Considering all possible sequential information structures can be daunting, but we can restrict attention to some canonical structures without loss of generality. This result appears in analogous forms in Myerson [1986], Forges [1986], Kamenica and Gentzkow [2011], and Bergemann and Morris [2016], as the "Revelation Principle" or, as we will call it, the "Obedience Principle".

First, we define the subset of canonical sequential information structures. In what follows, let $\text{Id}_A$ refer to the identity mapping from $A$ to $A$.

**Definition 6.** $(\sigma, \pi, p)$ *is an* **obedient triple** *if $S = A$ and $\sigma = \text{Id}_A$.*

An obedient triple is given by a prior, an information structure that recommends an action, and a strategy of the agent that always obeys the recommendation. When an action sequence can be rationalized with an obedient triple, we say that it has an *obedient rationalization*. We can now state and prove the Obedience Principle.

**Lemma 2** (Obedience principle). *If $\mathbf{a}$ can be rationalized, then it has an obedient rationalization.*

*Proof.* Suppose that $\mathbf{a}$ is rationalized by $(\sigma, \pi, p)$. We show that $\mathbf{a}$ is also rationalized by $(\text{Id}_A, \sigma \circ \pi, p)$. First, note that, in the information structure $\sigma \circ \pi$, the set of signals is $A$, so strategies become deviation rules. Second,
$$\sigma \circ \pi \circ p\,(\mathbf{a}) = \text{Id}_A \circ (\sigma \circ \pi) \circ p\,(\mathbf{a})$$
by associativity of composition and $\text{Id}_A$ is adapted. Hence, if $\mathbf{a}$ is chosen with positive probability under $(\sigma, \pi, p)$, it also is under $(\text{Id}_A, \sigma \circ \pi, p)$. Now we must show that $\text{Id}_A$ will be optimal for $(\sigma \circ \pi, p)$, whenever $\sigma$ is optimal for $(\pi, p)$. Suppose that an alternate strategy $D : A \to \Delta(A)$ does better than $\text{Id}_A$ when facing $(\sigma \circ \pi, p)$. In terms of payoff, it is easy to check that $U\,(\text{Id}_A, \sigma \circ \pi, p) = U\,(\sigma, \pi, p)$ and $U\,(D, \sigma \circ \pi, p) = U\,(D \circ \sigma, \pi, p)$. So if $(D, \sigma \circ \pi, p)$ gives a higher expected payoff than $(\text{Id}_A, \sigma \circ \pi, p)$, then the deviation $(D \circ \sigma, \pi, p)$ gives a higher payoff than $(\sigma, \pi, p)$ as well, implying that $\sigma$ was not optimal. □

## 5.2 Polyhedral Switching Lemma

The standard min-max theorem used to prove the Wald-Pearce lemma in Section 3 cannot be directly applied, one of the reasons being that the argument under "max" is not a compact set. To circumvent this problem, we prove a result that helps us switch the order of quantifiers in the final step of the proof of Theorem 1.

**Lemma 3** (Polyhedral Switching Lemma). *Let $X \subset \mathbb{R}^m$ be defined by a finite system of linear inequalities (weak or strict) and let $Y \subset \mathbb{R}^n$ be a polytope.*[13] *If $f : \mathbb{R}^m \times \mathbb{R}^n \to \mathbb{R}$ is affine in each variable, then the following statements are equivalent:*

1. *For every $x \in X$, there exists $y \in Y$ such that $f\,(x, y) > 0$;*

2. *There exists $y \in Y$ such that, for every $x \in X$, $f\,(x, y) > 0$.*

In our setup, the hyperplane argument is simplified by the finite setting, and thus the problem reduces to a linear program, which is what Lemma 3 captures.

---

[13] A set defined by a finite number of (weak or strict) linear inequalities is called an "evenly convex polyhedron" (Fajardo et al. [2020]). A polytope is the convex hull of a finite number of points or, equivalently, a compact set defined by finitely many weak linear inequalities.

## 5.3 Proof of Theorem 1

*The "only if" direction: if **a** is truly dominated, it cannot be rationalized.* Let $D$ be a deviation rule that dominates **a**. We show that any strategy that plays **a** with positive probability cannot be optimal. Indeed, given an arbitrary $(\sigma, \pi, p)$, we can define an alternative strategy $\tilde{\sigma} = D \circ \sigma$. Now consider how the expected payoff of the agent changes by switching from $\sigma$ to $\tilde{\sigma}$. Let $\gamma$ denote the joint distribution over $(\mathbf{b}, \omega)$ which is induced by $(\sigma, \pi, p)$. The difference in payoffs then becomes

$$U(\tilde{\sigma}, \pi, p) - U(\sigma, \pi, p) = \mathbb{E}_\gamma [u(D(\mathbf{b}), \omega) - u(\mathbf{b}, \omega)]$$

For each $(\mathbf{b}, \omega)$, this difference is non-negative, with strict inequality for $\mathbf{b} = \mathbf{a}$. Hence, if $\gamma$ puts positive probability on **a**, the overall difference will be positive, meaning that the agent benefits strictly from deviating to $\tilde{\sigma}$. The exact inequalities that show $\tilde{\sigma}$ to be an improvement over $\sigma$ are presented in Claim 1 in the appendix.

*The "if" direction: if **a** cannot be rationalized, it is truly dominated.* Given an action sequence **a** which cannot be rationalized, we must find a deviation rule $D$ that dominates it. Letting $\Theta(\mathbf{a}) = \{(\sigma, p, \pi) | \sigma \circ \pi \circ p(\mathbf{a}) > 0\}$, we can write the statement "**a** cannot be rationalized" as

$$\forall\, (\sigma, \pi, p) \in \Theta(\mathbf{a})\ \exists\, \hat{\sigma} : S \to \Delta(A) \text{ adapted s.t. } U(\hat{\sigma}, \pi, p) > U(\sigma, \pi, p).$$

By the Obedience Principle (Lemma 2), the statement "**a** cannot be rationalized" is equivalent to the statement "**a** cannot be rationalized by an obedient triple". This means that we can, without loss, restrict attention to $\pi : \Omega \to \Delta(A)$ and to $\sigma = Id_A$ in the statement above. Moreover, given that the set of signals is now $A$, the set of all strategies $\hat{\sigma}$ is simply the set of all deviation rules, $\mathbb{D} = \{D : A \to \Delta(A) \text{ s.t. } D \text{ is adapted}\}$. Incorporating these, we get the equivalent statement

$$\forall\, p \in \Delta(\Omega)\ \&\ \pi : \Omega \to \Delta(A) \text{ s.t. } \pi \circ p(\mathbf{a}) > 0, \quad \exists\, D \in \mathbb{D} \text{ s.t. } U(D, \pi, p) > U(Id_A, \pi, p).$$

Our goal is to switch the order of quantifiers in this statement. Therefore, we invoke the Polyhedral Switching Lemma to do the needful. To use the lemma, we simplify the problem further.

First, the objective function can be made linear through a simple change of variables: Let $\gamma \in \Delta(A \times \Omega)$ be the joint distribution on $A \times \Omega$ induced by the pair $(\pi, p)$. That is, $\gamma(\mathbf{b}, \omega) = \pi(\mathbf{b}|\omega)p(\omega)$. Doing this, the objective function becomes

$$\mathbb{E}_\gamma [u(D(\mathbf{b}), \omega) - u(\mathbf{b}, \omega)],$$

which is bilinear in $(\gamma, D)$. The statement that **a** cannot be rationalized can be rewritten as

$$\forall\, \gamma \in \Delta(A \times \Omega) \text{ with } \gamma(\mathbf{a}) > 0, \exists\, D : A \to \Delta(A) \text{ s.t. } \mathbb{E}_\gamma [u(D(\mathbf{b}), \omega) - u(\mathbf{b}, \omega)] > 0.$$

Now let $X = \{\gamma \in \Delta(A \times \Omega) \text{ s.t. } \gamma(\mathbf{a}) > 0\}$, $Y = \mathbb{D}$, and $f \equiv \mathbb{E}_\gamma[u(D(\mathbf{b}), \omega) - u(\mathbf{b}, \omega)]$. Then $X$ is defined by a finite system of linear inequalities, $Y$ is a polytope and $f$ is an affine function in $\gamma$ and $D$ (see Claim 2 for details). Thus, the conditions of Lemma 3 are met, and we can flip the order of quantifiers to get:

$$\exists\, D : A \to \Delta(A) \text{ s.t. } \forall\, \gamma \in \Delta(A \times \Omega) \text{ with } \gamma(\mathbf{a}) > 0 : \ \mathbb{E}_\gamma[u(D(\mathbf{b}), \omega) - u(\mathbf{b}, \omega)] > 0.$$

Let $D^*$ be such a deviation rule. By construction, $D^*$ dominates $\mathbf{a}$, and thus $\mathbf{a}$ is truly dominated.

## 6 Rationalizing distributions

In the previous section, we characterized the empirical content of the model when the analyst observes a single action sequence. We now consider a situation where the analyst has information about a large population of agents, so that his data consists of an entire distribution of chosen action sequences. Two results using deviation rules are presented, which assume varying levels of data availability. We end the section with a comparison between these results and Theorem 1.

### 6.1 Distributions over actions and states

At first, we assume that the analyst has access to a rich dataset that records both action sequences and realized states. That is, the analyst observes an entire joint distribution $\gamma \in \Delta(A \times \Omega)$. The distributions that can result from the Bayesian model can then be defined by modifying Definition 3.

**Definition 7.** *A distribution $\gamma \in \Delta(A \times \Omega)$ can be* **rationalized** *if there exists $(\sigma, \pi, p)$ such that*

1. $\sigma \in \arg\max_{\hat{\sigma}} U(\hat{\sigma}, \pi, p)$, *and*

2. $\gamma(\mathbf{a}, \omega) = \sigma \circ \pi(\mathbf{a}|\omega) \cdot p(\omega) \ \forall\ \mathbf{a} \in A,\ \omega \in \Omega$.[14]

Thus, a joint distribution $\gamma$ is rationalized if there exists a sequential information structure and a prior such that, in best responding to them, the agent's optimal strategy generates $\gamma$. In the context of Example 1, we are assuming that a large number of firms face some prior and sequential information structure, and in best responding to it, a joint distribution over the two states and three possible action sequences is produced, which the analyst seeks to rationalize.

Our notion of dominance must take into account the distributional structure of information available to the analyst. Earlier, we required the deviation rule to improve upon every action sequence and state. Since the criterion of rationalization is now stronger, the notion of dominance must be weaker. Therefore, we look at the *average improvement* brought about by a deviation rule.

---

[14]Note that for a fixed $\gamma \in \Delta(A \times \Omega)$, the prior $p$ is necessarily its marginal on $\Omega$. This is implicit in part 2 of the definition. We keep the choice of the triplet $(\sigma, \pi, p)$ in the definition to maintain consistency with Definition 3.

**Definition 8.** *A deviation rule $D : A \to \Delta(A)$ **dominates** a distribution $\gamma \in \Delta(A \times \Omega)$ if*

$$\sum_{\mathbf{a},\omega} [u(D(\mathbf{a}),\omega) - u(\mathbf{a},\omega)] \gamma(\mathbf{a},\omega) > 0.$$

*We say that $\gamma$ is **dominated on average** if there exists a deviation rule that dominates it.*

To understand the appropriateness of this notion of dominance, we follow the Obedience Principle (Lemma 2): If $\gamma$ can be rationalized, it must be possible to find an obedient rationalization. Under such an obedient rationalization, any alternative strategy $\hat{\sigma}$ is an adapted mapping from $A$ to $A$, hence a deviation rule. [15] If, however, $\gamma$ cannot be rationalized, then the candidate obedient strategy is not optimal, and there must be an alternative strategy—a deviation rule—that improves upon it. This proves the following characterization.

**Theorem 2.** *A distribution $\gamma \in \Delta(A \times \Omega)$ cannot be rationalized if and only if it is dominated on average.*

*Proof. The "only if" direction: if $\gamma$ is dominated on average, it cannot be rationalized.* Let $D$ be a deviation rule that dominates $\gamma$. We show that any strategy that produces $\gamma$ as the joint distribution over $A \times \Omega$ cannot be optimal. Suppose $(\sigma, \pi, p)$ induces $\gamma$, and define an alternative strategy $\tilde{\sigma} = D \circ \sigma$. The difference in payoffs then becomes

$$U(\tilde{\sigma}, \pi, p) - U(\sigma, \pi, p) = \mathbb{E}_\gamma [u(D(\mathbf{b}),\omega) - u(\mathbf{b},\omega)].$$

which is positive since $D$ dominates $\gamma$. Hence $\gamma$ cannot be induced by an optimal strategy and cannot be rationalized.

*The "if" direction: if $\gamma$ cannot be rationalized, it is dominated on average.* Fix a $\gamma \in \Delta(A \times \Omega)$ which cannot be rationalized. We must find a deviation rule $D$ that shows that it is dominated on average. Letting $\Theta(\gamma) = \{(\sigma,\pi,p) | \gamma(\mathbf{a},\omega) = \sigma \circ \pi(\mathbf{a}|\omega) \cdot p(\omega)\}$, we can write the statement "$\gamma$ cannot be rationalized" as

$$\forall (\sigma,\pi,p) \in \Theta(\gamma) \ \exists \hat{\sigma} : S \to \Delta(A) \text{ adapted s.t. } U(\hat{\sigma},\pi,p) > U(\sigma,\pi,p).$$

Following the same steps as in the proof of Theorem 1, we use the obedience principle to rewrite the above statement as

$$\exists D : A \to \Delta(A) \quad \text{s.t.} \quad \mathbb{E}_\gamma [u(D(\mathbf{b}),\omega) - u(\mathbf{b},\omega)] > 0.$$

which shows that there exists a $D$ that dominates $\gamma$, and hence $\gamma$ is dominated on average.

$\square$

---

[15] For Theorem 2, it is without loss of generality to focus on pure deviation rules.

An equivalent way of thinking about Theorem 2 is this: a distribution $\gamma$ can be rationalized if, for all deviation rules $D : A \to \Delta(A)$,

$$\sum_{\mathbf{a},\omega} [u(\mathbf{a},\omega) - u(D(\mathbf{a}),\omega)] \gamma(\mathbf{a},\omega) \geq 0.^{16}$$

Now, fix $p$ to be the marginal of $\gamma$ on $\Omega$, $S = A$, and $\sigma = \mathrm{Id}_A$. Then, noting that $\gamma(\mathbf{a},\omega) = \pi(\mathbf{a}|\omega) p(\omega)$, the above inequality gives us a unique obedient triplet that rationalizes $\gamma$.

The result can be seen as a counterpart to obedience constraints in information design (see surveys by Bergemann and Morris [2019] and Kamenica [2019]). In the lexicon of that literature, all distributions $\gamma \in \Delta(A \times \Omega)$ that satisfy the above inequality for all deviation rules can be supported as "Bayes Correlated Equilibria" of our decision problem.

Theorem 2 also sheds light on how to find information structures that rationalize particular action sequences. An action sequence $\mathbf{a}$ can be rationalized if there exists a distribution $\gamma$ that can be rationalized and puts positive probability on $\mathbf{a}$. Any such distribution can be interpreted directly as an information structure that signals action recommendations. Since Theorem 2 characterizes all distributions that can be rationalized, we need only look at those that put positive probability on $\mathbf{a}$ to find all obedient triples that rationalize $\mathbf{a}$.

## 6.2 Distributions over actions

In some scenarios, observing the realized state of the world might be difficult or even impossible for the analyst. For instance, the analyst may observe the investment decisions made by the population of firms in Example 1, but may not observe whether the underlying market forces were good or bad for them. In that case, when can a given distribution over action sequences be rationalized?

Since a triple $(\sigma, \pi, p)$ defines a joint distribution $\gamma \in \Delta(A \times \Omega)$, we consider the set of such joint distributions that is consistent with a given marginal $\bar{\gamma} \in \Delta(A)$,

$$\Gamma(\bar{\gamma}) = \left\{ \gamma \in \Delta(A \times \Omega) \mid \sum_{\omega} \gamma(\mathbf{a},\omega) = \bar{\gamma}(\mathbf{a}) \right\}.$$

The definition of rationalizing a distribution of action sequences then corresponds to Definition 7, but where we allow any joint distribution with the given marginal $\bar{\gamma}$.

**Definition 9.** *A distribution $\bar{\gamma} \in \Delta(A)$ can be* **rationalized** *if there exists $(\sigma, \pi, p)$ such that*

1. $\sigma \in \arg\max_{\hat{\sigma}} U(\hat{\sigma}, \pi, p)$, and

2. $\sigma \circ \pi \circ p = \bar{\gamma}$.

To prove that $\bar{\gamma}$ cannot be rationalized, we must show that it is impossible to find any $\gamma \in \Gamma(\bar{\gamma})$ that can be rationalized in the sense of Definition 7. One way to do this is to find a deviation rule that works simultaneously for all distributions in $\Gamma(\overline{\gamma})$. This idea leads us to the concept of *intermediate domination*.

---

[16] It can be noted that the use of mixed deviation rules in this statement is redundant, if the set of inequalities hold for all pure deviation rules $D : A \to A$, the statement is still true.

**Definition 10.** *A deviation rule $D : A \to \Delta(A)$* **dominates** *a distribution $\overline{\gamma} \in \Delta(A)$ if*

$$\sum_{a} \min_{\omega} \left[ u\left(D\left(\mathbf{a}\right), \omega\right) - u\left(\mathbf{a}, \omega\right) \right] \overline{\gamma}\left(\mathbf{a}\right) > 0.$$

*We say that $\overline{\gamma}$ is* **intermediately dominated** *if there exists a deviation rule that dominates it.*

Thus $D$ dominates $\overline{\gamma}$ if the average improvement is positive, even when we choose the worst possible state for each action sequence. The requirement of Definition 10 is thus intermediate to the notions of true dominance, which looks at the worst improvement across states and actions, and average dominance, which looks at the average improvement across states and actions.

If $\overline{\gamma}$ is intermediately dominated, the deviation rule that dominates $\overline{\gamma}$ demonstrates that any $\gamma \in \Gamma(\overline{\gamma})$ cannot be rationalized, and thus $\overline{\gamma}$ cannot be rationalized. As the reader might suspect, the converse holds.

**Theorem 3.** *A distribution $\overline{\gamma} \in \Delta(A)$ cannot be rationalized if and only if it is intermediately dominated.*

*Proof.* As in the case of Theorem 1, one direction is easy. Suppose $D$ dominates $\overline{\gamma}$ in the sense of Definition 10, and, by contradiction, $\gamma$ can be rationalized by some triplet $(\sigma, \pi, p)$. Then, the alternative strategy $\hat{\sigma} = D \circ \sigma$ gives the agent a strictly higher expected payoff when facing $(\pi, p)$. Thus, it must be that $\overline{\gamma}$ cannot be rationalized.

Conversely, suppose that $\overline{\gamma}$ cannot be rationalized. Then, by Theorem 2, for every $\gamma \in \Delta(A \times \Omega)$ with marginal $\gamma_A = \overline{\gamma}$, there exists a deviation rule $D$ such that

$$\sum_{a,\omega} \left[ u\left(D\left(\mathbf{a}\right), \omega\right) - u\left(\mathbf{a}, \omega\right) \right] \gamma\left(\mathbf{a}, \omega\right) > 0.$$

By Lemma 3, we can invert the order of quantifiers: there exists a deviation rule $D$ such that for all $\gamma \in \Delta(A \times \Omega)$ with marginal $\gamma_A = \overline{\gamma}$, we have:

$$\sum_{a,\omega} \left[ u\left(D\left(\mathbf{a}\right), \omega\right) - u\left(\mathbf{a}, \omega\right) \right] \gamma\left(\omega|\mathbf{a}\right) \overline{\gamma}\left(\mathbf{a}\right) > 0.$$

Now, we are free to choose $\gamma(\omega|\mathbf{a})$ arbitrarily given $\mathbf{a}$, which means that the inequality above must hold for every $\gamma(\omega|\mathbf{a})$. This happens precisely when

$$\sum_{a} \min_{\omega} \left[ u\left(D\left(\mathbf{a}\right), \omega\right) - u\left(\mathbf{a}, \omega\right) \right] \overline{\gamma}\left(\mathbf{a}\right) > 0.$$

□

Noting the system of inequalities that defines the three notions of dominance can help understand the intermediate nature of this result. Condition (b) in Definition 5 implies that for true

dominance, we need

$$\min_{\mathbf{b} \in A} \min_{\omega \in \Omega} \left[ u\left(D\left(\mathbf{b}\right), \omega\right) - u\left(\mathbf{b}, \omega\right) \right] \geq 0$$

with a strict inequality for $\mathbf{b} = \mathbf{a}$, the particular action sequence being dominated. Intermediate dominance (Definition 10) modifies this identity by taking the average over the distribution of action sequences using the marginal $\bar{\gamma}$:

$$\sum_{\mathbf{b}} \min_{\omega} \left[ u\left(D\left(\mathbf{b}\right), \omega\right) - u\left(\mathbf{b}, \omega\right) \right] \bar{\gamma}\left(\mathbf{b}\right) > 0.$$

And, finally, average dominance (Definition 8) takes the average over both action sequences and states using the knowledge of the joint distribution $\gamma$:

$$\sum_{\mathbf{b}, \omega} \left[ u\left(D\left(\mathbf{b}\right), \omega\right) - u\left(\mathbf{b}, \omega\right) \right] \gamma\left(\mathbf{b}, \omega\right) > 0.$$

Now, we compare the three main results in terms of the empirical contents of the respective Bayesian models they characterize.

## 6.3 Discussion

The richer predictions afforded by Theorem 2 and 3 come at the backdrop of several assumptions on the environment that we now discuss.

Theorems 2 and 3 assumed that the analyst could observe a whole distribution of action sequences. What does that mean? A natural interpretation is that we have data on choices for a large population of agents, and the analyst observes the empirical distribution of choices for that population. Under that interpretation, we assumed that all agents have the same utility function, prior, and information structure. Moreover, we also need to assume that the signals seen by each agent are independent. Only then can we conclude, by a standard law of large numbers argument, that the empirical distribution should be close to the theoretical distribution generated by the triple $(\sigma, \pi, p)$. No such assumptions are required for Theorem 1.

The assumption that all agents have the same information structure can be relaxed without consequence. If we allow multiple information structures within a population, each information structure will generate a distribution that can be rationalized. The observed distribution in the population will then be a weighted average of those distributions, which can also be rationalized because the sets of distributions that can be rationalized in Theorems 2 and 3 are convex.[17]

The independence assumption, however, is quite important. Consider what would be the effect of relaxing it in Example 1. If any correlation is allowed, we could have that every agent faces a problem with the same relevant state of the world (G or B) and sees the same public signal. It is perfectly plausible that the public signal be first good and then bad, leading all agents in the population to choose the action sequence (*invest*, *pull back*). But then we could observe 100% of the population choose an apparently dominated action sequence, which would seem to violate Theorem 3. Without independence, all we can say is that the observed distribution must have its support on action sequences that are not truly dominated, highlighting the relevance of Theorem 1 even when population data is available.

---

[17]This observation was made in Gualdani and Sinha [2024] for the static information design problem with a fixed prior.

## 7 Using linear programs to find deviation rules

Theorems 1, 2, and 3 all require us to find a deviation rule to show that some data cannot be rationalized. But how can we find suitable deviation rules that establish dominance? Here we show that the search for a deviation rule is equivalent to solving a linear program.[18]

### 7.1 Deviation Rules are Linear Inequalities

First, note that the set of all deviation rules is defined by linear inequalities: It is the set of all $\mathbb{R}^{A \times A}$ such that two properties are satisfied:

1. *Probability density:*

$$\sum_{\mathbf{b}} D(\mathbf{b}|\mathbf{a}) = 1 \quad \forall \mathbf{a} \in A, \text{ and } D(\mathbf{b}|\mathbf{a}) \geq 0 \quad \forall \mathbf{a}, \mathbf{b} \in A$$

2. *Adaptedness:*

$$\sum_{b_{t+1},\ldots,b_T} D(\mathbf{b}|a_1, \ldots, a_t, a_{t+1}, \ldots a_T) = \sum_{b_{t+1},\ldots,b_T} D(\mathbf{b}|a_1, \ldots, a_t, a'_{t+1}, \ldots a'_T) \; \forall \mathbf{a}, \mathbf{a}', \mathbf{b} \in A$$

Thus, when we write "$D$ is a deviation rule", this means that $D \in \mathbb{R}^{A \times A}$ satisfies the linear inequalities above. We now present the equivalent linear programs for the three main results, from the simplest to the most complex program.

### 7.2 A linear program for Theorem 2

Theorem 2 provides a characterization of when a joint distribution over action sequences and states of the world cannot be rationalized, using the notion of domination on average. We can think of the problem of finding a deviation rule in this case as a problem of maximizing the expected gain from deviating. It follows immediately from Theorem 2 that the distribution cannot be rationalized if and only if this maximum gain from deviating is positive. We summarize this in the following proposition:

---

[18]We are grateful to two anonymous referees and the editor for encouraging us to formally explore the connection.

**Proposition 2.** $\gamma \in \Delta(A \times \Omega)$ *is dominated on average if and only if the following linear program has a value greater than zero:*

$$\max_{D \in \mathbb{R}^{A \times A}} \sum_{\mathbf{a},\mathbf{b},\omega} \left[ u(\mathbf{b},\omega) - u(\mathbf{a},\omega) \right] D(\mathbf{b}|\mathbf{a}) \gamma(\mathbf{a},\omega)$$

$$\text{s.t. } D \text{ is a deviation rule.}$$

*In that case, the solution $D$ to the linear program is a deviation rule that dominates $\gamma$.*

## 7.3 A linear program for Theorem 3

Theorem 3 provides a characterization of when a distribution over action sequences $\bar{\gamma} \in \Delta(A)$ cannot be rationalized, using the notion of intermediate dominance. Because of the minimization over $\omega$ in the characterization, the equivalent linear program is not as immediate, but can still be achieved by introducing a new variable $k \in \mathbb{R}^A$.

**Proposition 3.** $\bar{\gamma} \in \Delta(A)$ *is intermediately dominated if and only if the following linear program has a value greater than zero:*

$$\max_{D \in \mathbb{R}^{A \times A}, k \in \mathbb{R}^A} \sum_{\mathbf{a}} k_{\mathbf{a}} \bar{\gamma}(\mathbf{a})$$

$$\text{s.t. } D \text{ is a deviation rule}$$

$$k_{\mathbf{a}} \leq \sum_{\mathbf{b}} \left[ u(\mathbf{b},\omega) - u(\mathbf{a},\omega) \right] D(\mathbf{b}|\mathbf{a}) \qquad \forall \mathbf{a} \in A \text{ and } \omega \in \Omega$$

*In that case, the solution $D$ to the linear program is a deviation rule that dominates $\gamma$.*

*Proof.* Since $\bar{\gamma}(\mathbf{a}) \geq 0$, the maximization is solved by choosing, for each $\mathbf{a}$, the highest possible value of $k_{\mathbf{a}}$ satisfying the constraints

$$k_{\mathbf{a}} \leq \sum_{\mathbf{b}} \left[ u(\mathbf{b},\omega) - u(\mathbf{a},\omega) \right] D(\mathbf{b}|\mathbf{a}) \quad \forall \omega \in \Omega.$$

This is achieved by setting

$$k_{\mathbf{a}} = \min_{\omega \in \Omega} \sum_{\mathbf{b}} \left[ u(\mathbf{b},\omega) - u(\mathbf{a},\omega) \right] D(\mathbf{b}|\mathbf{a}).$$

Thus, the linear program in the statement has the same value and solution $D$ as the (non-linear) program

$$\max_{D \in \mathbb{R}^{A \times A}} \sum_{\mathbf{a}} \min_{\omega \in \Omega} \sum_{\mathbf{b}} \left[ u(\mathbf{b},\omega) - u(\mathbf{a},\omega) \right] D(\mathbf{b}|\mathbf{a}) \bar{\gamma}(\mathbf{a})$$

$$\text{s.t. } D \text{ is a deviation rule}$$

which has a positive value if and only if its solution $D$ dominates $\overline{\gamma}$. □

### 7.4 A linear program for Theorem 1

The most permissive of the results, requiring minimal data, Theorem 1, assumes the outside analyst observes a single realized action sequence and asks whether it can be rationalized. In doing so, it relies on the notion of true dominance through deviation rules. The following result shows that the search for deviation rules can once again be formalized through a suitably constructed, but distinct, linear program.

**Proposition 4.** $\mathbf{a} \in A$ *is truly dominated if and only if the following linear program has a value greater than zero:*

$$\max_{D \in \mathbb{R}^{A \times A}, k \in \mathbb{R}} k$$

s.t. $D$ is a deviation rule

$$\sum_{\mathbf{c}} \left[ u(\mathbf{c}, \omega) - u(\mathbf{b}, \omega) \right] D(\mathbf{c}|\mathbf{b}) \geq 0 \qquad \forall \mathbf{b} \in A \text{ and } \omega \in \Omega$$

$$k \leq \sum_{\mathbf{c}} \left[ u(\mathbf{c}, \omega) - u(\mathbf{a}, \omega) \right] D(\mathbf{c}|\mathbf{a}) \qquad \forall \omega \in \Omega$$

*In that case, the solution $D$ to the linear program is a deviation rule that dominates $\mathbf{a}$.*

*Proof.* If there is a solution $D, k$ with $k > 0$, this means that $D$ satisfies

$$\sum_{\mathbf{c}} \left[ u(\mathbf{c}, \omega) - u(\mathbf{b}, \omega) \right] D(\mathbf{c}|\mathbf{b}) \geq 0 \qquad \forall \mathbf{b} \in A \text{ and } \omega \in \Omega$$

$$0 < k \leq \sum_{\mathbf{c}} \left[ u(\mathbf{c}, \omega) - u(\mathbf{a}, \omega) \right] D(\mathbf{c}|\mathbf{a}) \qquad \forall \omega \in \Omega$$

which means that $D$ dominates $\mathbf{a}$. On the other hand, if $D$ truly dominates $\mathbf{a}$, we can set

$$k = \min_{\omega \in \Omega} \sum_{\mathbf{c}} \left[ u(\mathbf{c}, \omega) - u(\mathbf{a}, \omega) \right] D(\mathbf{c}|\mathbf{a}) > 0$$

and we will have that $D$ and $k$ satisfies all of the restrictions. This means that the value of the linear program is at least $k$, so it must be greater than zero. □

## 8 Applications and connections to the literature

We now present four applications and use these to connect the paper to varied literatures in economic theory. First, Corollary 2 shows that when the agent is more risk averse, more action sequences can be rationalized. Second, we argue how Theorem 3 can be deployed to reject the standard Bayesian model simply using data on sequential action choices made by a population of agents. Third, we argue how the characterization of empirical content through Theorems 2 and 3 connects us to a burgeoning literature in dynamic information design. Fourth, we show how deviation rules, through Theorems 1 and 3, can be used to partially identify preference parameters without assumptions on beliefs or information. We end the section with a further discussion of related work.

## 8.1 Impact of risk aversion

The set of action sequences that can be rationalized is inextricably connected with the agent's utility function. It may be possible to ask how the set of action sequences that can be rationalized changes as the utility function of the agent is changed systematically. Here, we show that the set of actions that can be rationalized increases with risk aversion. Thus, if we can rule out an action sequence for an agent with a utility function $u$, we can also rule out that action sequence for all agents with a utility function $v$ which is less risk averse than $u$.

Recall that $v$ is less risk averse than $u$ if and only if there exists an increasing and convex function $f : \mathbb{R} \to \mathbb{R}$ such that $v = f \circ u$. Using this fact, Weinstein [2016] and Battigalli, Cerreia-Vioglio, Maccheroni, and Marinacci [2016] show that the set of rationalizable strategies increases with risk aversion. Using Theorem 1, the same logic can be applied here.

**Corollary 2.** *Let $v, u : A \times \Omega \to \mathbb{R}$ be two utility functions, with $v$ less risk averse than $u$. If $\mathbf{a}$ cannot be rationalized for $u$, then it cannot be rationalized for $v$.*

*Proof.* Let $f$ be an increasing convex function such that $v = f \circ u$. By Theorem 1, $\mathbf{a}$ cannot be rationalized for $u$ if and only if there exists a deviation rule $D : A \to \Delta(A)$ such that $u(D(\mathbf{b}), \omega) \geq u(\mathbf{b}, \omega)$ for all $\mathbf{b}$ and $\omega$, with a strict inequality for $\mathbf{b} = \mathbf{a}$. The same deviation rule will work for $v$, since, by Jensen's inequality,

$$
\begin{aligned}
v(D(\mathbf{b}), \omega) &= \sum_{\mathbf{c} \in A} f \circ u(\mathbf{c}, \omega) D(\mathbf{c}|\mathbf{b}) \\
&\geq f\left[\sum_{\mathbf{c} \in A} u(\mathbf{c}, \omega) D(\mathbf{c}|\mathbf{b})\right] \\
&= f(u(D(\mathbf{b}), \omega)) \\
&\geq f(u(\mathbf{b}, \omega)) \\
&= v(\mathbf{b}, \omega)
\end{aligned}
$$

The first inequality follows from Jensen's inequality, and the second follows from the definition of true dominance and is strict for $\mathbf{b} = \mathbf{a}$. Hence $\mathbf{a}$ is truly dominated for $v$. □

The corollary is stated purely in terms of what actions can be rationalized. But the proof is in terms of deviation rules, and we are not aware of any proof that would work directly in the information space, hence the relevance of Theorem 1 in proving the result.

## 8.2 Rejecting the 'standard' model

If we have data on a population of agents and see many of them taking an apparently dominated action sequence, we may question whether this behavior is consistent with our Bayesian model. What is the maximum probability of taking that action sequence that can still be rationalized? Since the action sequence is apparently dominated, we know that this upper bound must be less than one. If the fraction of agents taking this action sequence exceeds the upper bound, then the Bayesian model, under the appropriate assumptions discussed in Section 6.3, is rejected.

Suppose the highest probability is $\gamma \in [0,1]$.[19] First, a lower bound is obtained for $\gamma$ by constructing a specific information structure, and then an upper bound is obtained by devising a specific deviation rule. An educated guess for each side makes these bounds coincide, leading to a precise value for $\gamma$. We illustrate this for the two examples from the introduction.

**Revisiting Example 1.** We know that all three action sequences can be rationalized. Since (*invest, pull back*) is apparently dominated, it can be rationalized at most with some probability $\gamma \in (0,1)$. For the lower bound on $\gamma$, suppose we start with a uniform prior: both good and bad states are equally likely. Consider a sequential information structure that gives no information in the first period, and in the second period gives information according to the following conditional probability system:

|      | g | b |
|------|---|---|
| good | $\alpha$ | $1-\alpha$ |
| bad  | 0 | 1 |

When the state is good, signal $g$ is generated with probability $\alpha$ and signal $b$ is generated with probability $1-\alpha$, and when the state is bad, signal $b$ is generated for sure. Thus, conditional on investing in the first period, the agent will choose to continue investing in the second period if the signal is $g$. It can be checked by applying Bayes' rule that the agent will pull back upon seeing signal $b$ if and only if $\alpha \geq \frac{2}{3}$.[20] So, assume $\alpha \geq \frac{2}{3}$.

The agent's expected payoff if he chooses to invest in the first period is given by:

$$\frac{1}{2} \cdot (-1) + \frac{1}{2} \cdot [2\alpha + (1-\alpha)(-1)] = \frac{3}{2}\alpha - 1$$

Clearly, it is optimal to invest in the first period if $\alpha \geq \frac{2}{3}$. Finally, the probability with which the agent will choose (*invest, pull back*) under this information structure is

$$\frac{1}{2} \cdot 1 + \frac{1}{2} \cdot (1-\alpha),$$

---

[19]Note that $\gamma = 0$ for a truly dominated action sequence (Theorem 1), and it is easy to see that it equals 1 for an action sequence that is not apparently dominated. The interesting case, that is $gamma \in (0,1)$, arises when the action sequence is apparently dominated but not truly dominated.

[20]Note $\mathbb{P}(\text{good}|b) = \frac{1-\alpha}{2-\alpha}$. It is optimal to choose to pull back if $2 \cdot \mathbb{P}(\text{good}|b) + (-2) \cdot \mathbb{P}(\text{bad}|b) \leq -1$, which is the case when $\alpha \geq \frac{2}{3}$.

which, given the constraints on $\alpha$, is maximized at $\alpha = \frac{2}{3}$. Thus, the distribution generated over the three action sequences $\{(\textit{not invest}, \emptyset), (\textit{invest}, \textit{pull back}), (\textit{invest}, \textit{invest})\}$ with $\alpha = \frac{2}{3}$ is $(0, \hat{\gamma}, 1 - \hat{\gamma})$ where $\hat{\gamma} = \frac{2}{3}$. We can conclude that $\gamma \geq \frac{2}{3}$.

Next, we characterize the upper bound on $\gamma$ using deviation rules. Without loss of generality, consider a distribution of the form $(0, \hat{\gamma}, 1 - \hat{\gamma})$, and consider the deviation rule described in the following table:

|   | *not invest* | *invest* & *pull back* | *invest* & *invest* |
|---|---|---|---|
| $D$ | *not invest* | *not invest* | *not invest* |

The expression in Theorem 3 is given by:

$$\hat{\gamma} \cdot 1 + (1 - \hat{\gamma}) \cdot (-2) > 0 \quad \text{that is} \quad 3\hat{\gamma} - 2 > 0.^{21}$$

Thus, any $\hat{\gamma} \geq \frac{2}{3}$ is "dominated" by this deviation rule which means that $\gamma \leq \hat{\gamma} \leq \frac{2}{3}$.

Collectively, we can conclude that the highest probability with which the action sequence (*invest*, *pull back*) can be rationalized in Example 1 is $\gamma = \frac{2}{3}$.

**Revisiting Example 2.** We are interested in the question, what is the maximum probability with which $wx$ can be rationalized as a function of $\delta$? Call this number $\gamma_\delta$. As before, we first construct a lower bound using a suitable information structure and then an upper bound using deviation rules, and choose wisely so that these coincide.

The calculations are a bit more involved, for we want to report the probability as an arbitrary function of $\delta$. It is immediate from previous discussions that $\gamma_\delta = 0$ for $\delta < \frac{4}{5}$, since $wx$ can only be rationalized for $\delta \geq \frac{4}{5}$. It is also clear that $\gamma_\delta < 1$ for $\delta < 1$, and it exactly equal to 1 for $\delta = 1$. In the appendix, we show that in fact:

$$\gamma_\delta = \left(3 - \frac{2}{\delta}\right) \cdot \mathbb{1}\left(\delta \geq \frac{4}{5}\right),$$

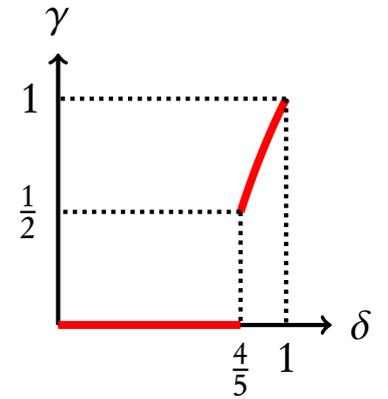

where $\mathbb{1}$ is the indicator function. The adjoining figure plots this value of $\gamma_\delta$. The construction of the information structure and the deviation rules that delivers $\left(3 - \frac{2}{\delta}\right)$ as the lower and upper bounds, respectively, are provided in Section 10.3.

## 8.3 Dynamic information design

Information design asks whether some desirable behavior can be induced with a careful choice of information structure, starting from a fixed prior belief on states. The results in this paper allow us to ask a more permissive version of this question: what are the set of feasible action sequences (Theorem 1) or distribution over action sequences (Theorems 2 and 3) for any possible prior. This can be interpreted as a situation in which the designer controls not only the information structure, but also the distribution over states.

---

[21] It is clear from this expression why it is without loss of generality to consider a distribution of the form $(0, \hat{\gamma}, 1 - \hat{\gamma})$. For any positive weight on (*not invest*, $\emptyset$), we can move the mass to the other two actions sequences, since the deviation maps (*not invest*, $\emptyset$) to itself and we are evaluating the highest probability with which (*invest*, *pull back*) can be rationalized.

To illustrate the ideas here, we will use a two-period version of the "moving the goalposts" model in Ely and Szydlowski [2020].

**Example 3.** A manager wants the employee to put in effort in both periods of a two-stage project. Each round of effort costs $c$ to the manager. If the task is easy, one round of effort is sufficient for the desired output of $R$ for the agent, and if the task is hard, then two rounds of effort are required. Payoffs are presented in the Figure below.

|  | no effort | effort & no effort | effort & effort |
|---|---|---|---|
| hard | 0 | $-c$ | $R - 2c$ |
| easy | 0 | $R - c$ | $R - 2c$ |

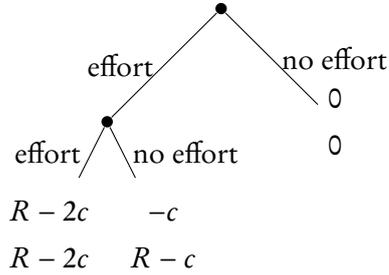

Assume $0 < c < R < 2c$. Then, the structure here is analogous to Example 1—it is a stopping problem, where one of the actions from continuing in the first period is apparently dominated by the choice to stop in the first period. The tools in this paper allow us to characterize the empirical content of the model—in particular, they tell us the highest probability with which (effort, effort) can be rationalized.

The highest probability that the action sequence (*effort*, *effort*) can be rationalized is $\frac{R-c}{c}$. This can be calculated similarly to how we found the highest possible probability for (*invest*, *pull back*) in Example 1 (see Section 8.2). This entails constructing a deviation rule that gives an upper bound on the probability and a simple information structure that gives the lower bound, which coincide. As a corollary, we can rationalize any fraction of agents from 0 to $\frac{R-c}{c}$ putting effort in both periods, without knowing any more information about how they learnt whether the task is easy or hard.

It seems only natural to ask for more: can deviation rules be similarly used to characterize the feasible distributions for a fixed prior, i.e., what class of distributions $\bar{\gamma}_p$ can be rationalized as a function of the prior $p$? Since this requires developing a new result, we refer the reader to our companion paper, de Oliveira and Lamba [2025a]. There, we present a result along the lines of Theorem 3 that works for a fixed prior; the idea of deviation rules is central there too. Applied to Example 3, this directly speaks to the model studied by Ely and Szydlowski [2020]. In recent work, Rehbeck [2023] and Doval et al. [2024] characterize similar feasibility results for the static model, when $T = 1$.

More broadly, dynamic information design is now a burgeoning area of study; see, for instance, Ely [2017], Doval and Ely [2020], Orlov, Skrzypacz, and Zryumov [2020], Basak and Zhou [2020]), and Makris and Renou [2023]. This, of course, builds on the static information design literature (Kamenica and Gentzkow [2011], Bergemann and Morris [2016]).

## 8.4 Partial identification of preferences

Suppose that the utility function $u$ belongs to a parametric class $\{u(\cdot, \lambda)\}_{\lambda \in \Lambda}$ and denote the true parameter by $\lambda_0$. We assume that the agent knows $\lambda_0$, but the analyst does not. The set of action sequences that can be rationalized would then depend on $\lambda$. Our goal here is to describe how to partially identify the parameter $\lambda_0$ using data on the distribution of actions chosen. In this exercise, we would like to know what values of $\lambda$ could be consistent with $\overline{\gamma}$, for some $p$ and $\pi$—in econometrics terms, we treat $p$ and $\pi$ as nuisance parameters. Thus, we ask: if the analyst observed the distribution over action sequences $\overline{\gamma} \in \Delta(A)$, what values of $\lambda$ would be consistent with such a choice, for some prior $p$ and information structure $\pi$?

Let $\Gamma(\lambda) \subseteq \Delta(A)$ denote the set of distributions that can be rationalized for a given value of $\lambda$. Upon observing the distribution $\overline{\gamma}$, the analyst would deduce that the $\lambda_0$ must lie in the sharp identified set

$$\Lambda(\overline{\gamma}) = \{\lambda \mid \overline{\gamma} \in \Gamma(\lambda)\}.$$

We can use deviation rules to approach this set as follows. Given a deviation rule $D$, let $\Lambda_D(\overline{\gamma})$ denote the set of $\lambda$'s for which $D$ does *not* dominate $\overline{\gamma}$ (in the sense of Definition 10). $\Lambda_D(\overline{\gamma})$ is typically easier to characterize than $\Lambda(\overline{\gamma})$ and, with the right choice of $D$, can provide substantial information. Moreover, the following result follows immediately from Theorem 3.

**Corollary 3.** $\Lambda(\overline{\gamma}) \subset \Lambda_D(\overline{\gamma})$ *for any deviation rule $D$, and* $\Lambda(\overline{\gamma}) = \bigcap_D \Lambda_D(\overline{\gamma})$.

Analogous methods also work for the notions of rationalization for Theorems 1 or 2. For Theorem 1, for instance, we would define $A(\lambda)$ as the set of action sequences that can be rationalized for a given value of $\lambda$ and the sharp identified set would be $\Lambda(\mathbf{a}) = \{\lambda \mid \mathbf{a} \in A(\lambda)\}$. This method was used in Example 2 to show that, upon observing that the agent chose to wait, the multiplicative waiting cost could not be greater than $\frac{4}{5}$; hence $\Lambda(wx) = \left[\frac{4}{5}, 1\right]$. We directly identified the set by constructing the "binding" deviation rule that recommends randomizing 50-50 in the first period instead of waiting. Similarly, by the argument in Section 8.2, upon observing a distribution $\overline{\gamma}$ that puts weight $\gamma^w > 0$ on $wx$ or $wy$, the analyst can conclude that $\delta \geq \max\{\frac{2}{3-\gamma^w}, \frac{4}{5}\}$; hence, we have $\Lambda(\overline{\gamma}) = \left[\max\{\frac{2}{3-\gamma^w}, \frac{4}{5}\}, 1\right]$.

Such partial identification exercises have been applied in static settings, using the notion of Bayes Correlated Equilibrium. Syrgkanis, Tamer, and Ziani [2021] estimate the distribution of valuations in an auction; Magnolfi and Roncoroni [2023] measure the effect of competition in an entry game; Gualdani and Sinha [2024] estimate a spatial voting model. These estimates have the advantage of making few assumptions on the information structure, which is typically hard to ascertain in practice. Our results show that such estimation exercises can potentially be extended to dynamic problems using the methods developed here.

## 8.5 Other literature

The most closely related papers to ours are the following: At the conceptual core, as pointed out in Section 3, the idea of using duality to characterize the empirical content of the dynamic decision problem builds on the static counterpart pioneered by Wald [1949] and Pearce [1984]. Further, in a static setting, Caplin and Martin [2015] provides a necessary condition for stochastic choices to be rationalized by information in a Bayesian model. The condition, called *no improving action switches*, states that no systematic reassignment of actions can lead to a higher expected utility. No improving action switches is analogous to true dominance, with the difference that our deviation rules include the adaptedness condition in order to respect the sequentiality of the problem. Also, Shmaya and Yariv [2016] explores the empirical implications of the Bayesian assumption in an experimentally motivated setting, providing a benchmark "anything goes" result for rationalization in dynamic choice. In contrast, we find that with richer settings but more limited data requirements, the sequential decision model can indeed make predictions.

Then, in closely related recent work on dynamic decision problems, a small theoretical literature seeks to rationalize the empirical content of dynamic decision problems: Taubinsky and Strack [2022] studies the question of identifying hyperbolic preference parameters from distributional choice data in a two-period problem. It argues that the time-consistent model can generally not be rejected—the only data that rejects it is the extreme case of complete preference reversal. On the other hand, Deb and Renou [2021] characterizes the empirical content of a model with multiple agents and common sequential learning. Even with limited data, it produces non-trivial predictions in applications for discrimination and committee voting.

There is a long tradition in economics of recovering parameters from observed choices, most notably in the revealed preference literature (see Chambers and Echenique [2016]). While most of the literature focuses on identifying utility functions, we take them, or at least the family they belong to, as given (though subject to the two subtleties mentioned in Section 2), and ask when choices can be explained via information. Information is also modeled as a general dynamic stochastic process in Chambers and Lambert [2021], where the agent also takes a sequence of actions. Their focus, however, is on finding the right utility function to elicit the overall information structure.

In the (axiomatic) decision theory literature, papers often seek to identify utilities and information simultaneously (see, for example, Dillenberger, Lleras, Sadowski, and Takeoka [2014], Piermont, Takeoka, and Teper [2016], and Lu [2016]). However, identifying this rich space of parameters requires much richer data as well, such as all choices from all menus. Frick, Iijima, and Strzalecki [2019] is an example that is close to our model. They study a dynamic random utility model, with one possibility being that the agent learns about their utility over time.

The question of which action sequences can be rationalized can also be expressed in terms of communication equilibria (see Myerson [1986] and Forges [1986], and more recently Sugaya and Wolitzky [2018]). The reformulated question becomes: what action sequences can occur with positive probability in a communication equilibrium of a single-player game?[22] Under this interpretation, our Obedience Principle (Lemma 2) is a particular case of the revelation principle of Myerson and Forges (see Propositions 1, 2, and 3 in Sugaya and Wolitzky [2018]), though our restricted context allows for a simpler proof. Myerson introduced the notion of codominated actions, which also extends the notion of a dominated action in a static multiplayer game to a multi-stage game. Although it seems reasonable to conjecture that the codomination procedure would eliminate all truly dominated action sequences under generic payoffs[23], in general, it only gives a sufficient condition for true dominance—there may be actions that are not codominated, but are never chosen together with positive probability in any communication equilibrium. For example, in Figure 3b, no actions are codominated, but the sequence of actions Lr is truly dominated.[24]

Further, in the context of Myersonian mechanism design, Rahman [2024] shows that Rochet [1987]'s characterization of incentive compatibility can be simplified to a set of detectable deviations and adapted to include both static multidimensional and dynamic problems. The idea of deviations there and the constructive proof through a zero-sum game between the principal and agents parallels Theorem 1 here.

## 9 Final remarks

This paper characterizes the empirical content of a standard Bayesian model for a general dynamic decision problem. Theorems 1, 2, and 3 vary the data requirement for the analyst and produce different implications for the Bayesian model, but they all rely on the same unifying feature: the idea of deviation rule, invoked each time we define dominance.

Theorem 1 gives the most parsimonious test of the model, by asking when an action sequence can be chosen by an optimizing agent. In contrast, Theorem 3 assumes that an entire distribution over action sequences can be observed, whereas Theorem 2 assumes further that states can also be observed. Put this way, it might seem that the results can be ranked, with Theorem 1 giving

---

[22]The embedding of our model within Myerson's multi-stage game works as follows: First, we introduce a player, called "Nature", who has constant payoffs and in the first period selects a state of the world $\omega \in \Omega$. The agent then selects actions from period 2 onward, with the final payoff to the agent given by $u(\mathbf{a}, \omega)$. In a communication equilibrium, Nature chooses $\omega$ according to a mixed strategy $p \in \Delta(\omega)$, reports this choice to the mediator, who then signals action recommendations to the agent.

In Myerson's framework, the agent can report their actions to the mediator, meaning that these reports could affect the information the agent receives. However, the agent cannot be punished for lying, since Nature only acts in period 1 and the mediator has no means of detecting a lie other than through the agent's own report. This means that, to understand the on-path distributions that result from communication equilibria, it is without loss of generality to restrict attention to information structures that do not depend on action reports, meaning that we can write the recommendations of the mediator as an action-independent information structure $\pi : \Omega \to \Delta(A)$

An alternative embedding can be obtained following Makris and Renou [2023], who consider a mediator who directly knows the state of the world.

[23]By generic, we mean that no two action sequences can give the same payoffs in the same state. This is a strong restriction, violated by many applications of interest.

[24]The key thing to understand is that Myerson's codominance is done period-by-period. So we first ask: having chosen L and being recommended to choose r, can the agent always strictly improve their payoff by deviating to $\ell$? The answer is no. Then we ask: in the first period, being recommended to choose $L$, can the agent strictly gain from deviating? No, because $L\ell$ is a best response to one state. This means that neither $L$ nor $r$ is codominated.

the weakest predictions, Theorem 3 intermediary, and Theorem 2 the strongest. However, these stronger predictions also come with stronger assumptions on the data-generating process, e.g., that the signals in the population are generated independently. Because of these accompanying assumptions, we see no obvious way to rank the theorems, leaving to the analyst the task of choosing the appropriate result.

It is natural to ask how hard it is to search the family of deviation rules. In Section 7, we show how each of the main theorems can be reduced to a linear program. In practice, the family of "relevant" deviation rules is drastically smaller than the set as we define them. In related work (see de Oliveira and Lamba [2025b]), we show how to reduce the complexity of what deviation rules would be relevant in establishing dominance. For example, in the context of Example 1, a total of twenty-seven pure deviation rules are possible, but only three of them are "relevant". More generally, for optimal stopping problems with two actions in each period, the set of potentially binding deviation rules in Theorem 1 can be reduced from being potentially exponential in $T$ to being exactly $T$.

Even if the analyst cannot run through the entire family of deviation rules, a single deviation rule can already lend empirical content to the dynamic decision problem. In contrast, constructing an information structure can only show that some distribution is possible but does not rule out any distribution. As we showed in the examples, instead of working through all possible deviation rules or all possible information structures, our results help to completely characterize the empirical content of the model through a few deviation rules and one information structure. Identifying systematically the binding deviation rules and information structures is a promising question for future work.

Finally, a reasonable question on counterfactuals goes as follows: What distributions of action sequences can be rationalized in a different decision problem, given the same information process? In the static model, Bergemann et al. [2022] show that the Bayes correlated equilibrium concept (BCE) from Bergemann and Morris [2016] can be used to derive such counterfactual predictions. We conjecture that an analogous construction can be done in our environment; a fuller exploration of this idea would make for interesting future work.

## 10 Appendix

### 10.1 Proof of Lemma 3

*Statement*: Let $X \subset \mathbb{R}^m$ be an evenly convex polyhedron and $Y \subset \mathbb{R}^n$ be a polytope (both non-empty). If $f : \mathbb{R}^m \times \mathbb{R}^n \to \mathbb{R}$ is affine in each variable, then the following statements are equivalent:

1. For every $x \in X$, there exists a $y \in Y$ such that $f(x, y) > 0$;

2. There exists a $y \in Y$ such that, for every $x \in X$, $f(x, y) > 0$.

*Proof.* (2) $\Rightarrow$ (1) is obvious, since we can just pick, for every $x$, the same $y$ that is given in (2).

We now prove (1) ⇒ (2). Since $Y$ is a polytope, it is the convex hull of its extreme points: $Y = conv(y_1, \ldots, y_J)$. Each $j$ determines an affine function $f(x, y_j)$. Thus, there is a $J \times m$ matrix $G$ and a $J \times 1$ vector $g$ such that we can write $f(x, y_j) = G_j x - g_j$ for every $j$.

Now, since $X$ is an evenly convex polyhedron, it is defined by a finite number of linear inequalities. Write those as $Ax \leq a$ and $Bx < b$. We can thus rewrite statement (1) as

1. There is no solution to the system of inequalities

$$Ax \leq a \quad Bx < b \quad Gx \leq g.$$

By Motzkin's Transposition Principle (Motzkin [1936]), this system does not have a solution if and only if there exist vectors $\alpha, \beta, \gamma \geq 0$ satisfying at least one of the two systems:

(a) $\alpha A + \beta B + \gamma G = 0$ and $\alpha a + \beta b + \gamma g < 0$

(b) $\alpha A + \beta B + \gamma G = 0$ and $\alpha a + \beta b + \gamma g \leq 0$ and $\beta \neq 0$.

Notice that if $\gamma = 0$, any of these two systems gives us a contradiction to the system of inequalities $Ax \leq a$ and $Bx < b$. Since we assume that $X$ is non-empty, we must have $\gamma \neq 0$. We may normalize $\gamma$ and the other variables so that $\sum_j \gamma_j = 1$. Now take any $x \in X$. From system (a), we get

$$\gamma G x = -\alpha A x - \beta B x \geq -\alpha a - \beta b > \gamma g$$

while from system (b) we get

$$\gamma G x = -\alpha A x - \beta B x > -\alpha a - \beta b \geq \gamma g.$$

Either way, we have that $\gamma (Gx - g) > 0$ for every $x \in X$. This finishes the proof, as can be seen by writing out the expression in its original terms: Letting $y = \sum_j \gamma_j y_j$, we have

$$f(x, y) = \sum_j \gamma_j (G_j x - g_j) = \gamma (Gx - g) > 0$$

for any $x \in X$. □

## 10.2 Completing the proof of Theorem 1

In the proof of Theorem 1, there are two missing pieces in Section 5.3, which we complete here. The first is the claim that completes the "only if" direction of the proof, and the second fits the final step that flips the order of quantifiers to Lemma 3.

**Claim 1.** *Fix $(\pi, p)$. If $\mathbf{a}$ is truly dominated by $D$ and strategy $\sigma$ is such that $\sum_{\omega \in \Omega} \sigma \circ \pi(\mathbf{a}|\omega) p(\omega) > 0$, then the strategy $\tilde{\sigma} = D \circ \sigma$ provides a strictly higher expected payoff.*

*Proof.* We show that the agent strictly benefits from switching from the strategy $\sigma$ to the strategy $\tilde{\sigma}$:

$$\begin{aligned}
U(\sigma, \pi, p) &= \sum_{\mathbf{b} \in A \setminus \{\mathbf{a}\}} \sum_{\omega \in \Omega} u(\mathbf{b}, \omega) \, \sigma \circ \pi(\mathbf{b}|\omega) \, p(\omega) \\
&\quad + \sum_{\omega \in \Omega} u(\mathbf{a}, \omega) \, \sigma \circ \pi(\mathbf{a}|\omega) \, p(\omega) \\
&< \sum_{\mathbf{b} \in A \setminus \{\mathbf{a}\}} \sum_{\omega \in \Omega} u(D(\mathbf{b}), \omega) \, \sigma \circ \pi(\mathbf{b}|\omega) \, p(\omega) \\
&\quad + \sum_{\omega \in \Omega} u(D(\mathbf{a}), \omega) \, \sigma \circ \pi(\mathbf{a}|\omega) \, p(\omega) \\
&= \sum_{\mathbf{b} \in A} \sum_{\omega \in \Omega} u(D(\mathbf{b}), \omega) \, \sigma \circ \pi(\mathbf{b}|\omega) \, p(\omega) \\
&= \sum_{\mathbf{b} \in A} \sum_{\omega \in \Omega} u(\mathbf{b}, \omega) \, D \circ \sigma \circ \pi(\mathbf{b}|\omega) \, p(\omega) \\
&= \sum_{\mathbf{b} \in A} \sum_{\omega \in \Omega} u(\mathbf{b}, \omega) \, \tilde{\sigma} \circ \pi(\mathbf{b}|\omega) \, p(\omega) \\
&= U(\tilde{\sigma}, \pi, p)
\end{aligned}$$

Note that the inequality above follows from Definition 5 (i.e., true dominance), and the strictness of it follows from part 1 of the definition and the fact that $\sum_{\omega \in \Omega} \sigma \circ \pi(\mathbf{a}|\omega) p(\omega) > 0$. □

**Claim 2.** *Let $X = \{\gamma \in \Delta(A \times \Omega) \text{ s.t. } \gamma(\mathbf{a}) > 0\}$, $Y = \{D : A \to \Delta(A) \text{ s.t. } D \text{ is adapted}\}$, and define $f : X \times Y \to \mathbb{R}$ as $f(\gamma, D) = \mathbb{E}_\gamma[u(D(\mathbf{b}), \omega) - u(\mathbf{b}, \omega)]$. Then,*

1. *$X$ is an evenly convex polyhedron;*

2. *$Y$ is a polytope; and*

3. *$f$ is an affine function in $\gamma$ and $D$.*

*Proof.* 1. This follows from the definition of evenly convex polyhedra, since $X$ is defined by a finite number of inequalities, namely

   (a) $0 \leqslant \gamma(\mathbf{a}, \omega) \leqslant 1$;
   
   (b) $\sum_{\mathbf{a}, \omega} \gamma(\mathbf{a}, \omega) = 1$; and
   
   (c) $\sum_\omega \gamma(\mathbf{a}, \omega) > 0$.

2. Here we consider $Y$ as a subset of $\mathbb{R}^{A \times A}$. Note that $Y$ is defined by a finite number of inequalities, namely

   (a) $0 \leqslant D(\mathbf{b}|\mathbf{a}) \leqslant 1$ for all $\mathbf{a}, \mathbf{b} \in A$;
   
   (b) $\sum_{\mathbf{b} \in A} D(\mathbf{b}|\mathbf{a}) = 1$ for all $\mathbf{a} \in A$; and

(c) The adaptedness restrictions, namely

$$\sum_{b_{t+1},\ldots,b_T} D(b_1,\ldots,b_t,b_{t+1},\ldots b_T | a_1,\ldots,a_t,a_{t+1},\ldots,a_T) =$$

$$\sum_{b_{t+1},\ldots,b_T} D(b_1,\ldots,b_t,b_{t+1},\ldots b_T | a_1,\ldots,a_t,c_{t+1},\ldots,c_T)$$

for all $a_1,\ldots a_T, b_1,\ldots b_t$, and $c_{t+1},\ldots,c_T$. Hence $Y$ is a polyhedron. It is also bounded, as can be seen from the inequalities in (a). Therefore, by Theorem 1.1 in Ziegler [2012], $Y$ is a polytope.

3. This follows immediately if we expand the definition of $f$:

$$f(\gamma, D) = \sum_{\mathbf{a},\mathbf{b},\omega} [u(\mathbf{a},\omega) - u(\mathbf{b},\omega)] D(\mathbf{a}|\mathbf{b})\gamma(\mathbf{b},\omega).$$

□

## 10.3 Solving for the highest probability of $wx$ in Example 2

We find the bound in two steps: construct a lower bound through information structures and an upper bound through deviation rules. Showing that these bounds coincide, the characterization is complete.

For a lower bound, start with a uniform prior, and consider a sequential information structure that gives no information in the first period, and in the second period, it gives information according to the following conditional probabilities:

|   | a | b |
|---|---|---|
| X | 1 | 0 |
| Y | $\alpha$ | $1-\alpha$ |

So when the agent sees signal $b$, she's sure that the state is $Y$; when she sees signal $a$, she puts some probability greater than $\frac{1}{2}$ on the state being $X$ (assuming $\alpha > 0$). Her optimal strategy is to choose $x$ after a signal of $a$ and to choose $y$ after a signal of $b$, with expected payoff of

$$\frac{1}{2}5\delta + \frac{1}{2}(\alpha 3\delta + (1-\alpha)\,5\delta) = 5\delta - \alpha\delta.$$

To be optimal to wait in the first period, we must have $5\delta - \alpha\delta \geq 4$, giving us the highest possible value of $\alpha$ to be $\alpha^* = 5 - \frac{4}{\delta}$. Thus, the probability that the agent chooses $wx$ under this information structure is the same as the probability that it results in a signal of $a$, namely

$$\frac{1}{2} + \frac{1}{2}\alpha^* = 3 - \frac{2}{\delta}.$$

To show that this is precisely the maximum probability that $wx$ can be chosen, we use Theorem 3. Consider the following two deviation rules: $D_x$, which takes $w$ to $x$ and $D_y$, which takes $w$ to $y$. Now, define $D_\lambda = \lambda D_x + (1-\lambda) D_y$, and write the gains from deviating in each state:

| deviation from wx | $u(D(wx), X) - u(wx, X)$ | $u(D(wx), Y) - u(wx, Y)$ |
|---|---|---|
| $D_x$ | $5 - 5\delta$ | $3 - 3\delta$ |
| $D_y$ | $3 - 5\delta$ | $5 - 3\delta$ |
| $D_\lambda$ | $\lambda(5 - 5\delta) + (1-\lambda)(3 - 5\delta)$ | $\lambda(3 - 3\delta) + (1-\lambda)(5 - 3\delta)$ |

| deviation from wy | $u(D(wy), X) - u(wy, X)$ | $u(D(wy), Y) - u(wy, Y)$ |
|---|---|---|
| $D_x$ | $5 - 3\delta$ | $3 - 5\delta$ |
| $D_y$ | $3 - 3\delta$ | $5 - 5\delta$ |
| $D_\lambda$ | $\lambda(5 - 3\delta) + (1-\lambda)(3 - 3\delta)$ | $\lambda(3 - 5\delta) + (1-\lambda)(5 - 5\delta)$ |

We would like to pick $\lambda$ so that the deviation is quite favorable when the agent is choosing $wx$, even in the worst-case state. So we choose $\lambda = \frac{1+\delta}{2}$ so that the payoffs under $wx$ are equated. Under that deviation rule, the expected worst-case benefit of deviating is

$$[4 - 4\delta] \gamma(wx) + [4 - 6\delta] \gamma(wy)$$

which is positive precisely when

$$\gamma(wx) > \left(3 - \frac{2}{\delta}\right)(\gamma(wx) + \gamma(wy)).$$

In particular, if $\gamma(wx) > 3 - \frac{2}{\delta}$, the agent would benefit from this deviation.